\documentclass[fleqn,usenatbib]{mnras}
\usepackage{fix-cm}
\usepackage{newtxtext,newtxmath}
\usepackage[T1]{fontenc}
\usepackage{adjustbox}

\DeclareRobustCommand{\VAN}[3]{#2}
\let\VANthebibliography\thebibliography
\def\thebibliography{\DeclareRobustCommand{\VAN}[3]{##3}\VANthebibliography}

\usepackage{graphicx}	% Including figure files
\usepackage{amsmath}	% Advanced maths commands
\usepackage{booktabs}
\usepackage{float}
\usepackage{multirow}

\title[BLR scales in IC~5287 and Mrk~845]
{The 2MIG isolated AGNs -- 4. Linking hot-dust reverberation to
spectroscopically predicted BLR scales in IC~5287 and Mrk~845}

\author[I.O. Izviekova et al.]{
I.O. Izviekova,$^{1,2}$\thanks{E-mail: izviekova@gmail.com}
I.B. Vavilova,$^{1}$
O.V. Kompaniiets,$^{1}$
O. Zamora,$^{3,4}$
R. Clavero$^{3,4}$
\\
$^{1}$Main Astronomical Observatory of the NAS of Ukraine, 27, Akademik Zabolotny St., Kyiv, 03143, Ukraine\\
$^{2}$International Centre for Astronomical, Medical and Ecological Research (ICAMER), 27, Akademik Zabolotny St., Kyiv, 03143, Ukraine\\
$^{3}$Instituto de Astrofísica de Canarias, 38205 La Laguna, Spain\\ 
$^{4}$Departamento de Astrofísica, Universidad de la Laguna, 38206 La Laguna, Tenerife, Spain
}

\date{Submitted to MNRAS, September 2026}

\pubyear{\the\year{}}

\begin{document}
\raggedbottom

\label{firstpage}
\pagerange{\pageref{firstpage}--\pageref{lastpage}}
\maketitle

% Abstract of the paper

\begin{abstract}
We investigate the relation between directly measured hot-dust reverberation scales and spectroscopically predicted broad-line region (BLR) scales in the isolated Seyfert galaxies IC~5287 and Mrk~845 (MCG+09-25-022), combining ZTF optical light curves, WISE/NEOWISE monitoring, and archival SDSS spectroscopy. Both nuclei show bluer-when-brighter optical variability, redder-when-brighter MIR behaviour, and delayed MIR responses consistent with circumnuclear dust reprocessing. For IC~5287, interpolated cross-correlation function analysis with flux-randomisation/random-subset-selection uncertainties yields a rest-frame $W1$ delay of $106\sb{-39}\sp{+24}$~d, corresponding to $0.089\sb{-0.033}\sp{+0.020}$~pc. The median $W2$ response occurs at a longer delay, but the lag distributions overlap substantially. The SDSS decomposition reveals strong host dilution and BLR reddening, $E(B-V)\sb{\rm BLR}=0.612\sb{-0.066}\sp{+0.067}$~mag. Applying the same colour excess to the AGN continuum as an illustrative scenario increases the predicted H$\beta$ BLR lag from approximately 4.8 to 11.7~d and reduces the dust-to-predicted-BLR radius ratios from approximately 20--25 to 8--10. For Mrk~845, transfer-function modelling gives a rest-frame $W1$ delay of $157\sb{-8}\sp{+19}$~d, corresponding to $0.132\sb{-0.007}\sp{+0.016}$~pc. Its $W2$ response occurs at a longer delay, with $P[\tau(W2)>\tau(W1)]=0.963$, supporting wavelength-dependent hot-dust structure. The SDSS spectrum shows more moderate BLR reddening, $E(B-V)\sb{\rm BLR}=0.197\sb{-0.039}\sp{+0.046}$~mag. After applying the wavelength--redshift correction appropriate for comparison with the WISE dust size--luminosity relations, both galaxies remain within the population scatter. Their comparatively large dust-to-predicted-BLR radius ratios therefore primarily reflect the spectroscopically predicted BLR scales. These results emphasise the need to distinguish directly measured dust reverberation radii from BLR radii inferred from single-epoch spectroscopy when interpreting the BLR--dust hierarchy.
\end{abstract}


\section{Introduction}
\label{sec:introduction}

The variable ultraviolet--optical continuum of an AGN heats the circumnuclear dust, which re-emits the absorbed radiation in the infrared after a light-travel-time delay. Measuring this delayed response provides a response-weighted characteristic radius of the dust-emitting region through dust reverberation mapping \citep{Barvainis1987,Suganuma2006, Koshida2014, Minezaki2019}. Long-baseline WISE/NEOWISE monitoring has extended dust reverberation
mapping to large AGN samples at 3--5~$\mu$m. \citet{Lyu2019}
recovered convincing $W1$ and $W2$ reverberation signals for a large
fraction of nearby quasars, while \citet{Yang2020} extended $W1$
dust-reverberation measurements to hundreds of quasars over a broad
redshift and luminosity range. Subsequent uniform analyses established
shallower-than-canonical $W1$ and $W2$ dust size--luminosity relations
and confirmed that the characteristic dust-response radius increases
with wavelength \citep{Chen2023,Mandal2024,Ayubinia2025}.
\citet{Ayubinia2025} further showed that the WISE dust-response scales
are systematically larger than directly measured H$\beta$ BLR scales,
providing the population context for the BLR--dust comparison
considered here.

The physical interpretation of the BLR--dust radial hierarchy depends on how the BLR scale is determined. Uniform reverberation-mapping analyses have shown that the H$\beta$ size--luminosity relation is shallower than the canonical $R_{\rm BLR}\propto L^{0.5}$ scaling and exhibits systematic offsets with accretion state \citep{WangWoo2024}. Building on a uniformly analysed reverberation-mapped sample, \citet{Woo2026} introduced a three-parameter relation linking the H$\beta$ lag to the nuclear optical luminosity and H$\beta$ line width. Reliable single-epoch BLR predictions therefore require accurate measurements of both quantities. A key distinction is whether both radial scales are measured directly by reverberation mapping or whether a directly measured dust radius is compared with a BLR radius inferred from single-epoch spectroscopy. Host-galaxy starlight can bias $L_{5100}$, particularly when the nuclear continuum has low contrast against the stellar emission \citep{Bentz2006, Bentz2013}, while internal attenuation can further suppress the observed continuum \citep{Bentz2023IC4329A, Gaskell2023Reddening}. Broad-line Balmer decrements provide a useful indicator of BLR reddening \citep{Dong2008, Lu2019Balmer}, although their interpretation also depends on accretion state, line responsivity, and radiative-transfer effects \citep{Wu2023Balmer, Son2025Balmer}. These factors propagate directly into the inferred $R_{\rm dust}/R_{\rm BLR,pred}$ hierarchy: they can modify the spectroscopically predicted BLR scale while leaving the independently measured dust reverberation radius unchanged.

Optical variability provides an independent view of the nuclear continuum that drives the dust response and can complement the single-epoch spectroscopic estimate of the AGN continuum. Bluer-when-brighter (BWB) behaviour is common in type-1 AGNs, although redder-when-brighter (RWB) and weak colour trends are also observed \citep{Sudan2025, Ojha2025}. In nearby AGNs, host-galaxy dilution can substantially modify the observed colour--brightness relation \citep{Sakata2010}, while flux--flux relations provide a complementary constraint on the colour of the correlated variable component \citep{Choloniewski1981, PozoNunez2012, Ramolla2015}. Season-resolved analyses can further test whether the stochastic optical variability contains recurrent characteristic structure on intermediate timescales \citep{Phillipson2020}. Together, these diagnostics help distinguish the observed, host-contaminated colour behaviour from the spectral properties of the variable nuclear component and provide an independent characterisation of the optical continuum whose variations are subsequently reprocessed by circumnuclear dust.

The host-galaxy environment provides a complementary context for interpreting nuclear and circumnuclear properties. The 2MASS Isolated Galaxies catalogue (2MIG) contains nearby systems selected to minimise the influence of comparable neighbours and strong ongoing tidal interactions \citep{Vavilova2009, Karachentseva2010}. Our multiwavelength studies have shown that 2MIG AGNs span a broad range of nuclear and host-galaxy properties \citep{Pulatova2015, Vavilova2015, Vavilova2016, Vasylenko2020, Kompaniiets2023} and inhabit a variety of large-scale environments \citep{Vavilova2021b, Kompaniiets2025a, Vavilova2026}. The 2MIG sample therefore provides a well-defined framework for examining nuclear variability and circumnuclear reprocessing in galaxies selected to have reduced influence from strong current external perturbations.

In the previous paper of this series, we measured optical--MIR dust-reverberation delays in the isolated NLSy1 galaxies Mrk~42 and Mrk~493 \citep{Izviekova2026Mrk42Mrk493}. Here we extend this time-domain framework to IC~5287 and Mrk~845, two isolated broad-line Seyfert galaxies, and combine the reverberation analysis with a detailed decomposition of their optical spectra. This combination allows the directly measured MIR dust-response scales to be compared with BLR scales predicted from host-corrected single-epoch spectroscopy, while explicitly examining the effects of host dilution, broad-line reddening, and the inferred nuclear continuum on the resulting BLR--dust radial hierarchy.

In this work, we use long-baseline ZTF $g$, $r$, and $i$ light curves together with WISE/NEOWISE $W1$ and $W2$ monitoring to determine the optical-to-MIR dust-reverberation scales and their wavelength dependence and to characterise the chromatic variability of both nuclei. We decompose archival SDSS spectra into stellar, AGN, Fe~II, and emission-line components to determine the host-corrected $L_{5100}$, broad- and narrow-line properties, internal reddening, and the spectroscopic quantities required to predict the H$\beta$ BLR scale. We then compare these predicted BLR scales with the independently measured dust-reverberation radii and assess how host dilution and internal attenuation propagate into the inferred $R_{\rm dust}/R_{\rm BLR,pred}$ hierarchy. As complementary variability diagnostics, we use season-resolved ZTF light curves to probe month-scale optical structure in Mrk~845 and dedicated IAC80 observations to test both nuclei for variability on hour-long timescales.

The paper is organised as follows. Section~\ref{sec:sample_data} describes the targets and observational data, and Section~\ref{sec:methods} presents the analysis methods. The results are presented in Section~\ref{sec:results} and discussed in Section~\ref{sec:discussion}. Section~\ref{sec:conclusions} summarises the main conclusions.

\section{Targets and observational data}
\label{sec:sample_data}

\subsection{Target galaxies}
\label{subsec:targets}
We analyse two nearby isolated Seyfert galaxies selected from the 2MIG isolated-AGN sample: IC~5287 (2MIG~3128) and Mrk~845 (MCG+09-25-022; 2MIG~2067). Their basic properties are summarised in Table~\ref{tab:basic_properties}. Both objects have archival SDSS spectroscopy and long-baseline ZTF and WISE/NEOWISE photometric coverage, enabling a combined spectroscopic and optical--MIR time-domain analysis.

\begin{table*}
\centering
\caption{Basic properties of the galaxies analysed in this work.}
\label{tab:basic_properties}
\renewcommand{\arraystretch}{1.15}
\setlength{\tabcolsep}{3.5pt}

\begin{tabular*}{\textwidth}
{@{\extracolsep{\fill}}lcccccccc@{}}
\hline
Object &
2MIG &
RA (J2000) &
Dec. (J2000) &
$z$ &
Morphology &
Activity$_{\rm lit}$ &
$i$ (deg) &
$\log(M_{\rm BH}/M_\odot)$ \\
\hline

IC~5287 &
2MIG 3128 &
23:09:20.28 &
+00:45:23.0 &
0.032399 &
$(R')$SB(r)b &
Sy~1.2 &
67.9 &
$7.48\pm0.06$ \\

Mrk~845 &
2MIG 2067 &
15:07:45.04 &
+51:27:09.6 &
0.0459164 &
SABb &
Sy~1 &
72.4 &
$7.43\pm0.49$ \\
\hline
\end{tabular*}
\vspace{2mm}
\begin{minipage}{\textwidth}
\textit{Notes.}
The 2MIG identifiers, coordinates, and literature morphological and nuclear classifications are based primarily on \citet{Pulatova2015}. The inclination angles of both galaxies are taken from HyperLEDA \citep{Makarov2014}. For IC~5287, the adopted morphology and Sy~1.2 classification are also consistent with \citet{Kompaniiets2025a}, which provides the adopted literature black-hole mass. For Mrk~845, the adopted black-hole mass is based on the stellar velocity-dispersion estimate compiled by \citet{Kompaniiets2025a}. The activity classes report the literature classifications; the spectral state at the SDSS epoch is determined independently from the decomposed emission lines.
\end{minipage}
\end{table*}

IC~5287 is a barred and ringed spiral hosting a broad-line Seyfert nucleus. Published optical classifications span Sy~1--1.2: \citet{Pulatova2015} listed it as an $(R)$SB(r)b Sy~1 galaxy, while spectroscopy obtained in 1996 was classified as Sy~1.2 by \citet{Pietsch1998} and subsequently included in the homogeneous analysis of \citet{Kollatschny2008}. A recent analysis of archival \textit{Chandra} data adopted the $(R')$SB(r)b/Sy~1.2 classification and inferred a relatively low accretion state, with $\log\lambda_{\rm Edd}\simeq-2.3$, together with $\log(M_{\rm BH}/M_\odot)=7.48\pm0.06$ \citep{Kompaniiets2025a}. The availability of historical broad-line spectroscopy, an independently estimated black-hole mass, and long-baseline optical and MIR monitoring makes IC~5287 well suited to a joint analysis of its nuclear spectral properties and dust-reverberation scale.

Mrk~845 is a highly inclined spiral hosting a broad-line Seyfert nucleus and has long been identified as an active galaxy in the Bo{\"o}tes void \citep{Kim2001BootesVoid}. Broad Balmer emission is present, while the SDSS DR9 line ratios reported by \citet{Pulatova2015} placed the source in the H~II region of the BPT diagnostics. In MaNGA, Mrk~845 (plate--IFU 8593--12705) was included among AGN candidates with a resolved extended narrow-line region \citep{Chen2019ENLR}. The galaxy also shows rapid, large-amplitude X-ray variability in ROSAT observations \citep{Kim2001BootesVoid}. The combination of broad-line emission, apparently discrepant optical diagnostic classifications, resolved narrow-line emission, and documented nuclear variability makes Mrk~845 a useful counterpart to IC~5287 for the combined spectroscopic and time-domain analysis performed here.

\subsection{ZTF and WISE/NEOWISE photometry}
\label{subsec:survey_photometry}

Long-term optical variability is characterised using public ZTF $g$, $r$, and $i$ photometry \citep{Bellm2019ZTF, Masci2019ZTF, Graham2019}. The long-baseline light curves are used to quantify the variability amplitude and chromatic behaviour, to provide the optical driver for the reverberation analysis, and, for Mrk~845, to examine seasonal structure on intermediate timescales.

The MIR data are obtained from WISE and its NEOWISE reactivation mission \citep{Wright2010WISE,Mainzer2014NEOWISE,Cutri2012WISE,Cutri2015NEOWISE}. We use the $W1$ and $W2$ bands, centred at 3.4 and 4.6~$\mu$m, respectively. The decade-long MIR light curves are sampled in observing windows typically separated by approximately six months and provide the long-term infrared variability used to measure the delayed response to the optical continuum.

The ZTF and WISE/NEOWISE survey photometry includes contributions from
both the variable nucleus and the host galaxy. The measurements thus
represent central-region photometry. Colour--magnitude relations
characterise the chromatic behaviour of this emission, while
flux--flux diagnostics constrain the colour of the correlated variable
component. The optical host contribution is quantified independently
from the SDSS spectral decomposition at the spectroscopic epoch. The
photometric filtering, temporal binning, colour matching,
seasonal-timescale analysis, and reverberation procedures are described
in Section~\ref{sec:methods}.
 
\subsection{SDSS optical spectroscopy}
\label{subsec:sdss_data}
 
Both galaxies have archival single-fibre spectra from the Sloan Digital Sky Survey (SDSS) \citep{York2000SDSS, Abdurrouf2022SDSSDR17}. The spectra sample the central regions through the 3-arcsec SDSS fibre and provide single-epoch optical spectroscopy substantially earlier than the ZTF and WISE/NEOWISE monitoring.

For IC~5287, we analyse SDSS plate--MJD--fibre 381--51811--455, obtained on 2000 September 24. The adopted redshift is $z=0.032399$. The observation consists of four 900-s integrations, giving a total exposure time of 3600~s. For Mrk~845, we use plate--MJD--fibre 1165--52703--231, obtained on 2003 March 5, with an adopted redshift of $z=0.0459164$ and a total exposure time of approximately 2500~s. For both spectra, we retain the calibrated flux density, inverse variance, native logarithmic wavelength sampling, and wavelength-dependent instrumental resolution.

The spectra are reanalysed to separate the stellar continuum, AGN continuum, optical Fe~II emission, and broad and narrow emission-line components. The decomposition provides host-corrected nuclear luminosities at 5100~\AA, broad-Balmer profile measurements and decrements, narrow-line diagnostic ratios, [O~III] kinematics, and stellar velocity dispersions. These quantities are used to characterise the nuclear optical spectrum, estimate broad- and narrow-line reddening, and predict the H$\beta$ BLR scale from single-epoch spectroscopy. The predicted BLR scales are subsequently compared with the independently measured MIR dust-reverberation radii.

The SDSS spectroscopy predates the ZTF and WISE/NEOWISE monitoring by many years. This temporal separation is explicitly considered when interpreting comparisons between the single-epoch spectroscopic luminosities, the later reverberation measurements, and luminosity-dependent scaling relations; no contemporaneous optical spectrum is available for either object. The detailed continuum, emission-line, stellar-kinematic, reddening, and BLR-scale analyses are described in Section~\ref{subsec:sdss_methods}.

\subsection{Dedicated IAC80 monitoring}
\label{subsec:iac80_data}

In addition to the long-term survey data, IC~5287 and Mrk~845 were observed with the 0.82-m IAC80 telescope at the Observatorio del Teide, Tenerife, to search for intraday optical variability. The observations were obtained in service mode with the CAMELOT2 imager. CAMELOT2 uses a $4096\times4112$ back-illuminated CCD with a pixel scale of $0.322^{\prime\prime}$ pixel$^{-1}$ and a useful field of view of approximately $11.8^{\prime}\times11.8^{\prime}$.

The IAC80 observations provide high-cadence optical time series on hour-long timescales that are not adequately sampled by the long-term ZTF monitoring. They are used to test whether either nucleus exhibits statistically significant intraday variability, thereby complementing the long-baseline optical variability analysis.

The observing dates, exposure times, monitoring durations, and numbers of exposures retained for the final analysis are listed in Table~\ref{tab:iac80_log}. The CCD reduction, differential aperture photometry, selection of comparison stars, and statistical assessment of intraday variability are described in Section~\ref{subsec:iac80_methods}, with additional observational and quality-control details given in Appendix~\ref{app:idv}. The comparison-star sequences, their APASS DR9 multiband photometry, and static finding charts are publicly available in the \textit{AGN Reference Fields} dataset \citep{Izviekova2026AGNfields}.

\section{Methods}
\label{sec:methods}

The long-term time-domain analysis largely follows the framework developed in our previous work \citep{Izviekova2026Mrk42Mrk493}, including the photometric processing, optical and MIR variability diagnostics, and dust-reverberation analysis. In the present study, this framework is extended by a season-resolved optical-timescale analysis of Mrk~845 and by a substantially expanded treatment of the optical spectroscopy. The archival SDSS spectra are decomposed into stellar, AGN, Fe~II, and emission-line components, with separate measurements of the broad- and narrow-line emission, stellar and ionised-gas kinematics, Balmer-decrement reddening, and the host-corrected nuclear continuum. These measurements provide the spectroscopic quantities required to predict the H$\beta$ BLR scale and to assess how host-galaxy dilution and broad-line kinematics affect the inferred BLR--dust radial hierarchy. The sensitivity of this hierarchy to internal attenuation is examined separately through an illustrative scenario based on the broad-line Balmer decrement.

\subsection{Survey photometry and variability diagnostics}
\label{subsec:phot_methods}

The ZTF and WISE/NEOWISE photometry was processed following the
quality-control, Galactic-extinction correction, and temporal-binning
procedures described in \citet{Izviekova2026Mrk42Mrk493}. ZTF
measurements were combined into nightly means, while individual
WISE/NEOWISE measurements were grouped into observing epochs. As
discussed in Section~\ref{subsec:survey_photometry}, the survey
measurements are treated as central-region photometry, which includes
both nuclear and host-galaxy emission.

Optical colour--magnitude relations were fitted using York regression, including the covariance introduced by the shared magnitude, while flux--flux relations were fitted with orthogonal-distance regression. For a relation of the form $$F_{\nu,1}=aF_{\nu,2}+b,$$ the slope $a$ determines the colour, and hence the spectral index, of the correlated variable component.

Spectral indices were calculated following the prescription adopted in our previous variability analyses \citep{Izviekova2026Mrk42Mrk493,Kompaniiets2026NGC3521}, using the convention $F_\nu\propto\nu^{-\alpha}$. Colour-derived spectral indices and their brightness dependence, ${\rm d}\alpha/{\rm d}(-m)$, were obtained from the effective wavelengths of the corresponding filters; WISE magnitudes were converted to the AB system for this calculation. The spectral index of the correlated variable component, $\alpha_{\rm var}$, was derived from the corresponding flux--flux slope using the same convention.

The optical trends were additionally tested with seeing thresholds of $3\arcsec$ and $2\arcsec$. The survey colours are analysed as observed central-region photometry, with the host and AGN contributions
quantified independently from the SDSS spectral decomposition.

\subsubsection{Seasonal optical-timescale analysis of Mrk~845}
\label{sec:seasonal_gls}

The ZTF $g$- and $r$-band light curves of Mrk~845 were divided into
seven observing seasons to test for recurrent optical structure on
intermediate timescales. After weighted linear detrending, a
generalised Lomb--Scargle (GLS) periodogram was calculated over
20--110~d for each season. The location of the strongest GLS peak
defines the dominant seasonal GLS timescale, $P_{\rm GLS}$, which
characterises the variability structure within each observing season.
All seasonal timescales and recurrence intervals are quoted in the
observed frame.

The significance of recurrent power within a common timescale interval was evaluated using 30,000 parametric Monte Carlo realisations generated from linear-trend plus damped-random-walk (DRW) models fitted to the individual seasonal light curves. Each simulated light curve was subjected to the same detrending, GLS, and recurrence-search procedure as the observed data. Recurrence statistics were evaluated separately for the $g$ and $r$ bands and jointly for the $g+r$ data. Full definitions of the recurrence statistics, robustness tests, and comparisons with alternative stochastic models are given in Appendix~\ref{app:mrk845_timescale_tests}.

\subsection{Optical--MIR reverberation analysis}
\label{subsec:lag_methods}
 
The ZTF $g$-band light curve was adopted as the optical driver. Before measuring the delays, the variable accretion-disc contribution to the WISE bands was estimated by extrapolating the optical continuum into the MIR as a power law, $F_\nu\propto\nu^{s_{\rm AD}}$, and subtracted following standard dust-reverberation prescriptions \citep{Koshida2014, Lyu2019} and the implementation adopted in our previous work \citep{Izviekova2026Mrk42Mrk493}. Here $s_{\rm AD}$ denotes the power-law exponent used specifically for the accretion-disc extrapolation and is distinct from the colour spectral index $\alpha$ defined in Section~\ref{subsec:phot_methods}. We adopt $s_{\rm AD}=1/3$ for the fiducial calculation, corresponding
to the long-wavelength asymptotic slope of a standard thin accretion
disc \citep{Koshida2014}. We repeat the analysis with
$s_{\rm AD}=0.1$ and $0$ as flatter-continuum sensitivity tests
\citep{Minezaki2019}.

For both galaxies, observed-frame delays were converted to the source
rest frame and to characteristic response-weighted dust radii as
\begin{equation}
\tau_{\rm rest}
=
\frac{\tau_{\rm obs}}{1+z},
\qquad
R_{\rm dust}
=
c\tau_{\rm rest}.
\label{eq:rdust}
\end{equation}
Here $R_{\rm dust}$ denotes the characteristic reverberation-response
scale traced by the corresponding MIR band.

For comparison with the WISE dust size--luminosity relations of
\citet{Ayubinia2025}, we additionally account for the redshift-induced
shift of the rest-frame wavelength sampled by a fixed observed WISE
band. Following their empirical wavelength dependence,
$\tau\propto\lambda^{0.74}$, the population-comparison quantities are
\begin{equation}
\tau_{\rm corr}
=
\tau_{\rm obs}(1+z)^{-0.26},
\qquad
R_{\rm dust,corr}
=
c\tau_{\rm corr}.
\label{eq:dust_redshift_corr}
\end{equation}
This correction combines cosmological time dilation with the
wavelength--redshift dependence and is applied only in comparisons
with the $R_{\rm dust}$--$L_{5100}$ relations of
\citet{Ayubinia2025}. The source-frame light-travel delays and radii
used for the reverberation measurements and BLR--dust radius ratios
remain those defined in equation~\ref{eq:rdust}.

The probability of wavelength ordering,
$P[\tau(W2)>\tau(W1)]$, was evaluated from the propagated $W1$ and
$W2$ lag distributions.

\subsubsection{IC~5287: ICCF analysis}
\label{subsec:methods_ic5287_lag}
 
For IC~5287, the disc-corrected, epoch-averaged $W1$ and $W2$ light curves were cross-correlated with the ZTF $g$-band driver using the bidirectional interpolated cross-correlation function (ICCF). The fiducial lag was defined as the centroid of the contiguous primary positive-lag peak satisfying
\begin{equation}
r(\tau)\geq0.8r_{\rm max}.
\end{equation}
Lag uncertainties were estimated from 5000
flux-randomisation/random-subset-selection (FR/RSS) realisations,
with the 16th, 50th, and 84th percentiles of the valid centroid
distribution defining the reported median and uncertainty interval.
Convergence was verified by comparing cumulative subsets of
1000--5000 realisations.

Robustness was tested against the assumed accretion-disc slope, ICCF centroid threshold, lag-search interval, and removal of individual WISE/NEOWISE epochs. The corresponding diagnostics are presented in Appendix~\ref{app:ic5287_phot_tests}.
 
\subsubsection{Mrk~845: transfer-function modelling}
\label{subsec:methods_mrk845_lag} 

For Mrk~845, the long-term MIR variability was modelled as a delayed and temporally smoothed response to the ZTF $g$-band driver,
\begin{equation}
F_{\rm MIR}(t)
=
C
+
A\int
\Psi(\tau)F_g(t-\tau)\,{\rm d}\tau .
\label{eq:tf_model}
\end{equation}
where $\Psi(\tau)$ is the transfer function, $A$ is the response amplitude, and $C$ is a constant MIR component.

Top-hat and Gaussian transfer functions were tested independently. For each response model, the characteristic lag $\tau$, response width, and intrinsic scatter $s_{\rm int}$ were fitted together with the linear amplitude and constant terms, $A$ and $C$. For an MIR measurement $F_i$ with uncertainty $\sigma_i$, we define
\begin{equation}
\sigma_{{\rm eff},i}^{2}=\sigma_i^2+s_{\rm int}^2
\end{equation}
and evaluate the Gaussian likelihood as
\begin{equation}
-2\ln\mathcal{L}
=
\sum_i
\left[
\frac{(F_i-F_{{\rm mod},i})^2}
{\sigma_{{\rm eff},i}^{2}}
+
\ln\left(2\pi\sigma_{{\rm eff},i}^{2}\right)
\right].
\end{equation}
Model preference was assessed using
\begin{equation}
{\rm BIC}
=
-2\ln\mathcal{L}_{\max}
+
k\ln N .
\end{equation}
where $N$ is the number of fitted MIR epochs and $k=5$ is the total number of fitted parameters. The linear parameters $A$ and $C$ were optimised conditionally for each set of transfer-function parameters but were retained in the BIC parameter count.

Statistical lag uncertainties were obtained from bootstrap realisations, while comparison of the top-hat and Gaussian solutions was used to assess the dependence on the assumed response-function shape. The characteristic delays are substantially more stable than the fitted response widths, which approach the allowed boundary in a significant fraction of the realisations. We therefore use the widths only to characterise the fitted response shape and do not interpret them as robust measurements of the geometrical radial thickness of the dust-emitting region.

Additional tests varied the accretion-disc subtraction and the treatment of long interpolation gaps and repeated the analysis after removal of individual WISE/NEOWISE epochs. The final Mrk~845 lag uncertainties were enlarged beyond the statistical bootstrap intervals to encompass the accepted robustness tests. Transfer-function profile maps, response-function comparisons, and the corresponding robustness tests are presented in Appendix~\ref{app:mrk845_lag_tests}.

\subsection{Optical spectral analysis}
\label{subsec:sdss_methods}
 
\subsubsection{SDSS spectra, continuum decomposition, and stellar kinematics}
\label{subsec:continuum_methods}
 
The archival SDSS spectra described in Section~\ref{subsec:sdss_data} were reanalysed on their native logarithmic wavelength grids, explicitly accounting for the wavelength-dependent instrumental resolution. The spectra were corrected for foreground Galactic extinction using the \citet{Cardelli1989} extinction law with $R_V=3.1$. The stellar and nuclear continua were decomposed with \textsc{pPXF} \citep{Cappellari2017,Cappellari2023}, using E-MILES stellar templates \citep{Vazdekis2016EMILES}, an AGN power law $F_{\lambda,\rm AGN}\propto\lambda^\beta$, and the optical Fe~II template of \citet{Park2022FeII}. Emission-line regions and other spectral features unsuitable for continuum fitting were masked.

Alternative fitting wavelength ranges, emission-line masks, and stellar-kinematic configurations were used to assess continuum-model systematics. Solutions in which nuisance parameters were driven to imposed fitting boundaries were excluded from the adopted systematic envelope, while the remaining accepted continuum solutions were propagated through the subsequent emission-line analysis.

The AGN fraction at 5100~\AA\ was defined as
\begin{equation}
f_{\rm AGN,5100}
=
\frac{F_{\lambda,\rm AGN}(5100)}
{F_{\lambda,\rm AGN}(5100)+F_{\lambda,\star}(5100)} .
\label{eq:fagn5100}
\end{equation}

For IC~5287, the statistical uncertainty in $f_{\rm AGN,5100}$ was
estimated from 150 pixel-noise Monte Carlo realisations of the
spectrum. In each realisation, the continuum decomposition was
refitted with the fiducial Fe~II broadening and velocity shift held
fixed, while the AGN power-law slope was re-optimised. The final quoted
interval combines the 16th--84th percentile Monte Carlo range with the
envelope of accepted continuum-model solutions, adopting the outer
lower and upper bounds as a conservative uncertainty interval.

The fiducial $L_{5100}$ used in the BLR and dust size--luminosity
comparisons is derived from the host-subtracted AGN continuum corrected
for foreground Galactic extinction, with internal attenuation treated
separately.

The stellar velocity dispersion was measured independently with \textsc{pPXF}, accounting for the wavelength-dependent SDSS line-spread function and matching the stellar templates to the instrumental resolution. Strong AGN emission-line regions were masked. Statistical uncertainties were estimated from Monte Carlo realisations, while alternative wavelength ranges and mask configurations were used to assess systematic effects. Additional host-contamination and stellar-kinematic tests are presented in Appendices~\ref{app:ic5287_line_diagnostics} and \ref{app:mrk845_spectral_tests}.
 
\subsubsection{Emission-line decomposition and internal reddening}
\label{subsec:line_methods}
 
After subtraction of the stellar, AGN, and Fe~II continua, the main Balmer and forbidden-line complexes were modelled with multi-component Gaussian profiles. Narrow Balmer and forbidden lines were tied to common narrow-line kinematics where appropriate, while the broad Balmer components were fitted independently. The [O~III] profile was tested with both a narrow-core model and a core-plus-wing model to allow for asymmetric ionised-gas kinematics.

For IC~5287, one- and two-Gaussian models were compared for the strongly non-Gaussian broad H$\beta$ profile using the BIC. Across the accepted profile realisations, the broad H$\beta$ profile can switch between connected and split half-maximum topologies, making the conventional FWHM unstable. We therefore adopt the flux-weighted line dispersion, $\sigma_{\rm line}({\rm H}\beta)$, as the primary velocity measure and additionally calculate non-parametric cumulative-flux widths. The full profile-topology analysis and width definitions are given in Appendix~\ref{app:ic5287_hbeta}.

For Mrk~845, one- and two-component broad-line models were tested for both H$\beta$ and H$\alpha$, while core-only and core-plus-wing models were compared for [O~III]. Model selection was based on the BIC. The H$\beta$ line dispersion is retained as the velocity measure used for the BLR-scale prediction; detailed model-selection results are presented in Appendix~\ref{app:mrk845_spectral_tests}.

Broad- and narrow-line Balmer decrements were measured separately. We adopt an intrinsic broad-line ratio $(\mathrm{H}\alpha/\mathrm{H}\beta)_0=3.06$ \citep{Dong2008} and an NLR ratio of 3.1 \citep{OsterbrockFerland2006}, with colour excesses calculated using the \citet{Cardelli1989} extinction curve.

The host-subtracted continuum corrected only for foreground Galactic extinction defines the fiducial nuclear luminosity. To assess the possible sensitivity of the inferred BLR scale to internal continuum attenuation, we additionally consider an illustrative scenario in which
\begin{equation}
E(B-V)_{\rm cont}
=
E(B-V)_{\rm BLR}.
\end{equation}
Under this assumption, the AGN continuum luminosity and power-law slope are recalculated. This scenario is not treated as an independent measurement of the continuum reddening, but as a sensitivity test of how a BLR-reddening-based attenuation correction would modify $L_{5100}$ and the predicted BLR scale. For IC~5287, the resulting continuum slope is also compared with the standard long-wavelength thin-disc expectation, $F_\lambda\propto\lambda^{-7/3}$; the corresponding tests are presented in Appendix~\ref{app:ic5287_reddening}.
 
\subsubsection{Narrow-line diagnostics and Seyfert subtype}
\label{subsec:classification_methods}
 
The decomposed narrow-line fluxes were used to construct the standard [N~II], [S~II], and, where available, [O~I] BPT diagnostic diagrams, using only the narrow Balmer components. The [S~II] $\lambda6716/\lambda6731$ ratio was used as a density-sensitive diagnostic, while the [O~III] profile was used to characterise asymmetric ionised-gas kinematics. For IC~5287, we additionally measured the optical Fe~II strength $R_{\rm FeII}$ over 4434--4684~\AA\ relative to broad H$\beta$.

The optical Seyfert subtype was determined from the decomposed emission lines following \citet{Winkler1992}. The complete line analysis was propagated through 1000 Monte Carlo realisations for IC~5287 and 300 realisations for Mrk~845, including line-fitting uncertainties, emission-line topology, BPT classification, and Seyfert subtype. Additional diagnostics are presented in Appendices~\ref{app:ic5287_line_diagnostics} and \ref{app:mrk845_spectral_tests}.

\begin{figure*}
\centering
\includegraphics[width=0.49\textwidth]
{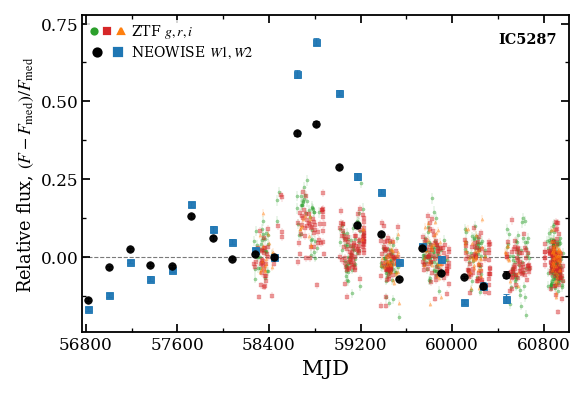}
\includegraphics[width=0.49\textwidth]
{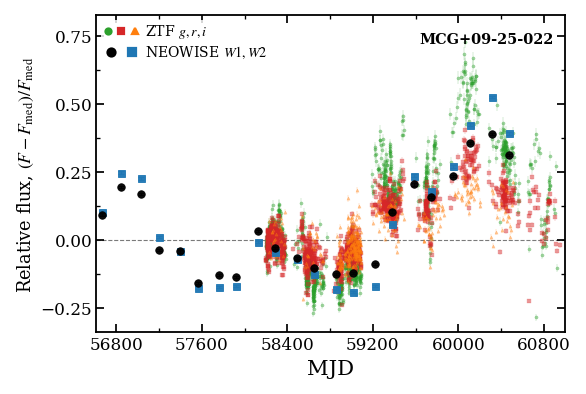}
\caption{Long-term optical and MIR variability of IC~5287 (left) and Mrk~845 (right). ZTF $g$, $r$, and $i$ measurements and epoch-averaged WISE/NEOWISE $W1$ and $W2$ measurements are shown relative to their median levels.}
\label{fig:longterm_both}
\end{figure*}
 
\subsubsection{Predicted H$\beta$ BLR scales}
\label{subsec:blr_prediction_methods}
 
The characteristic H$\beta$ BLR scale was predicted for both galaxies using the size--luminosity--velocity relation of \citet{Woo2026}. Our primary calculation adopts their mean-spectrum H$\beta$ line-dispersion calibration,
\begin{equation}
\log
\left(
\frac{\tau_{\mathrm{H}\beta}}{{\rm day}}
\right)
=
(0.47\pm0.05)\log L_{44}
+
(0.89\pm0.15)\log\sigma_3
+
(1.17\pm0.07).
\label{eq:woo_sigma_blr}
\end{equation}
where
\begin{equation*}
L_{44}
=
\frac{\lambda L_\lambda(5100\,{\rm \AA})}
{10^{44}\ {\rm erg\,s^{-1}}},
\qquad
\sigma_3
=
\frac{\sigma_{\rm line}({\rm H}\beta)}
{10^3\ {\rm km\,s^{-1}}}.
\end{equation*}

The intrinsic scatter of 0.21~dex in the adopted calibration, the calibration-coefficient uncertainties, the continuum-luminosity uncertainty, and the uncertainty in $\sigma_{\rm line}({\rm H}\beta)$ were propagated simultaneously. The calculation was performed for both the fiducial host-subtracted continuum and the illustrative BLR-reddening-based continuum-attenuation scenario.

The resulting $\tau_{\rm BLR,pred}$ values are single-epoch spectroscopic predictions from the adopted size--luminosity--velocity calibration and are not direct reverberation measurements of the BLR radius. The independently measured dust-response scales were compared with these predictions through
\begin{equation}
\frac{R_{\rm dust}}{R_{\rm BLR,pred}}
=
\frac{\tau_{\rm dust}}{\tau_{\rm BLR,pred}}.
\end{equation}
with both delays expressed in the source rest frame. For IC~5287, sensitivity to the conventional FWHM-based calibration and to the two FWHM topology branches is examined separately in Appendix~\ref{app:ic5287_reddening}.

\subsection{Intraday optical variability}
\label{subsec:iac80_methods}
 
Intraday variability was assessed using the power-enhanced $F$-test \citep{deDiego2014}, adopting a significance level of $\alpha_{\rm sig}=0.01$. For each observing sequence, the null hypothesis of no intrinsic target variability was rejected when the measured $F_{\rm enh}$ exceeded the corresponding critical value $F_{\rm crit}$ for the appropriate degrees of freedom. The same criterion was applied independently to IC~5287 and Mrk~845. Detailed observing, differential-photometry, comparison-star, and quality-control diagnostics are presented in Appendix~\ref{app:idv}.

\section{Results}
\label{sec:results}

The results are organised by physical diagnostic to facilitate a direct comparison between IC~5287 and Mrk~845. We first present the long-term chromatic variability of both nuclei and the season-resolved optical-timescale analysis of Mrk~845. We then report the optical--MIR reverberation scales and the results of the SDSS spectral decomposition, and finally compare the single-epoch spectroscopically predicted BLR scales with the independently measured hot-dust reverberation radii.
 
\subsection{Long-term variability and chromatic behaviour}
\label{subsec:results_variability}

\begin{table*}
\centering
\caption{Chromatic, spectral-index, and flux--flux diagnostics. $m$ is the York colour--magnitude slope, $r$ the Pearson coefficient, and $a$ the ODR flux--flux slope. The convention is $F_\nu\propto\nu^{-\alpha}$.}
\label{tab:chromatic_summary}
\small
\setlength{\tabcolsep}{2.6pt}
\renewcommand{\arraystretch}{1.08}
\begin{tabular*}{\textwidth}{@{\extracolsep{\fill}}llccccccc@{}}
\hline
Object &
Relation &
$m$ &
$r$ &
$\alpha_{\rm med}$ &
${\rm d}\alpha/{\rm d}(-X)$ &
$a$ &
$C_{\rm var}$ &
$\alpha_{\rm var}$ \\
\hline

IC~5287 &
$g-r$ vs. $g$ &
$0.630\pm0.054$ &
$0.638$ &
$2.879\pm0.014$ &
$-1.975\pm0.170$ &
$0.429\pm0.036$ &
$0.920$ &
$2.89$ \\

IC~5287 &
$r-i$ vs. $r$ &
$0.691\pm0.108$ &
$0.594$ &
$1.753\pm0.030$ &
$-3.079\pm0.483$ &
$0.767\pm0.123$ &
$0.288$ &
$1.28$ \\

IC~5287 &
$W1-W2$ vs. $W1$ &
$-0.519\pm0.096$ &
$-0.763$ &
$-0.902\pm0.056$ &
$+1.509\pm0.279$ &
$0.814\pm0.052$ &
$0.862$ &
$0.65$ \\

Mrk~845 &
$g-r$ vs. $g$ &
$0.479\pm0.009$ &
$0.935$ &
$2.173\pm0.015$ &
$-1.503\pm0.029$ &
$0.960\pm0.014$ &
$0.045$ &
$0.14$ \\

Mrk~845 &
$r-i$ vs. $r$ &
$0.329\pm0.027$ &
$0.672$ &
$1.353\pm0.018$ &
$-1.464\pm0.119$ &
$1.040\pm0.042$ &
$-0.043$ &
$-0.19$ \\

Mrk~845 &
$W1-W2$ vs. $W1$ &
$-0.263\pm0.039$ &
$-0.799$ &
$-0.627\pm0.033$ &
$+0.765\pm0.115$ &
$0.954\pm0.032$ &
$0.690$ &
$0.15$ \\
\hline
\end{tabular*}

\vspace{1mm}
\begin{minipage}{0.98\textwidth}
\footnotesize
For the optical relations, $X=g$ for $g-r$ and $X=r$ for $r-i$; for the MIR relation, $X=W1$. Optical colours and optical $C_{\rm var}$ values are Galactic-extinction corrected and expressed in the AB system. WISE colours and WISE $C_{\rm var}$ values are quoted in the Vega system. The colour--magnitude slopes are unchanged by the additive Vega-to-AB zero-point conversion. The WISE $\alpha_{\rm med}$ values are calculated from the corresponding AB colours, while $\alpha_{\rm var}$ is derived directly from the flux--flux slope. $\alpha_{\rm med}$ describes the observed central-region emission, whereas $\alpha_{\rm var}$ describes the correlated variable component.
\end{minipage}
\end{table*}

Both IC~5287 and Mrk~845 show clear long-term variability in the ZTF optical and WISE/NEOWISE MIR light curves (Fig.~\ref{fig:longterm_both}). In both galaxies, the optical variability amplitude increases towards shorter wavelengths. Their optical colour--magnitude relations show bluer-when-brighter (BWB) behaviour, whereas the MIR colour--magnitude relations show redder-when-brighter (RWB) behaviour.

For IC~5287, the central 90 per cent magnitude ranges are approximately 0.26, 0.22, and 0.19~mag in $g$, $r$, and $i$, respectively. The principal optical $g-r$ colour--magnitude relation has a York
slope of $m_{g-r}=0.630\pm0.054$ ($r=0.638$, $p<10^{-6}$), whereas the MIR relation gives ${\rm d}(W1-W2)/{\rm d}W1=-0.519\pm0.096$ ($r=-0.763$, $p=5.9\times10^{-5}$). In the adopted $F_\nu\propto\nu^{-\alpha}$ convention, the corresponding brightness dependences are ${\rm d}\alpha/{\rm d}(-g)=-1.975\pm0.170$ and ${\rm d}\alpha/{\rm d}(-W1)=+1.509\pm0.279$, respectively, quantifying the optical BWB and MIR RWB trends.

The principal flux--flux relations give $a_{gr}=0.429\pm0.036$ and $a_{W1W2}=0.814\pm0.052$. The corresponding variable-component spectral indices are $\alpha_{\rm var}=2.89$ in $g-r$ and $0.65$ in $W1-W2$ (Table~\ref{tab:chromatic_summary}); thus, the correlated optical variable component of IC~5287 remains relatively red in the adopted $F_\nu\propto\nu^{-\alpha}$ convention. The principal optical and MIR colour--magnitude and flux--flux relations are shown in Fig.~\ref{fig:ic5287_colours}, with additional optical colour combinations and flux--flux diagnostics presented in Appendix~\ref{app:ic5287_phot_tests}.

\begin{figure*}
\centering
\includegraphics[width=0.48\textwidth]
{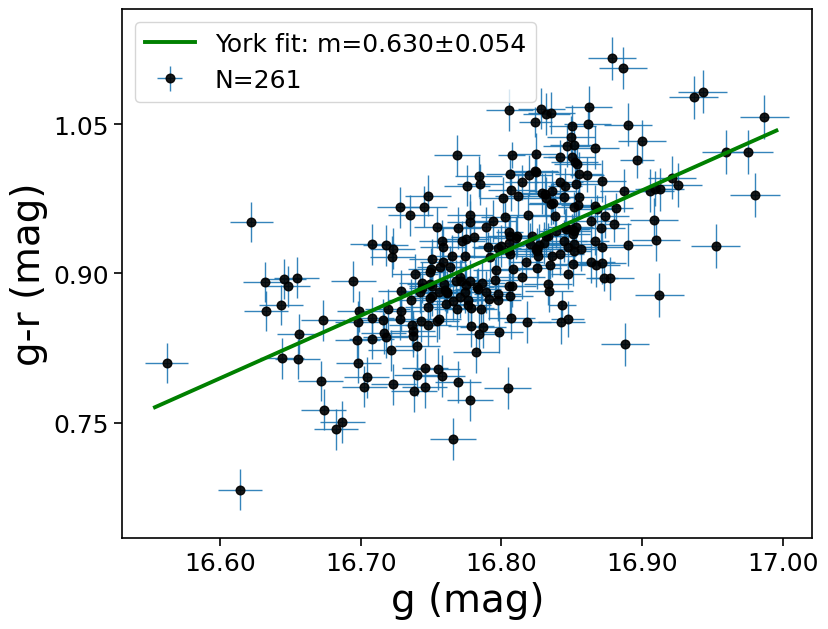}
\includegraphics[width=0.48\textwidth]
{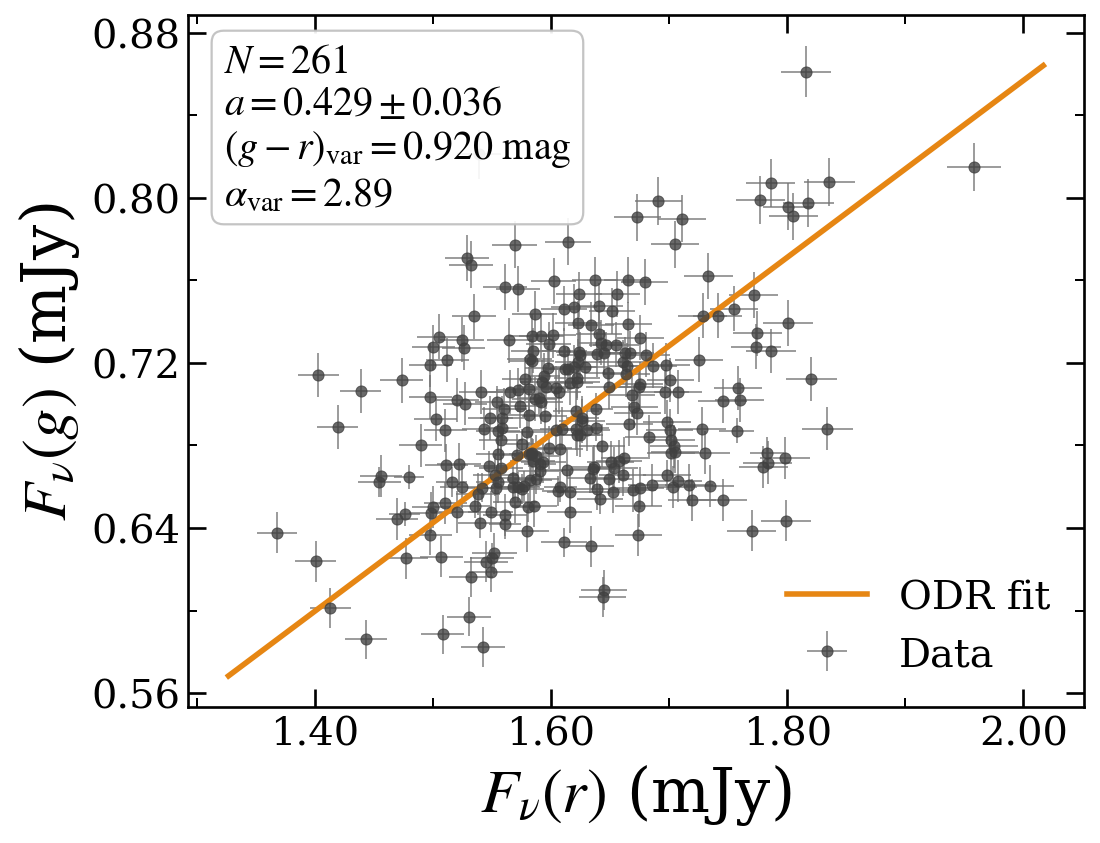}
\vspace{2mm}
\includegraphics[width=0.48\textwidth]
{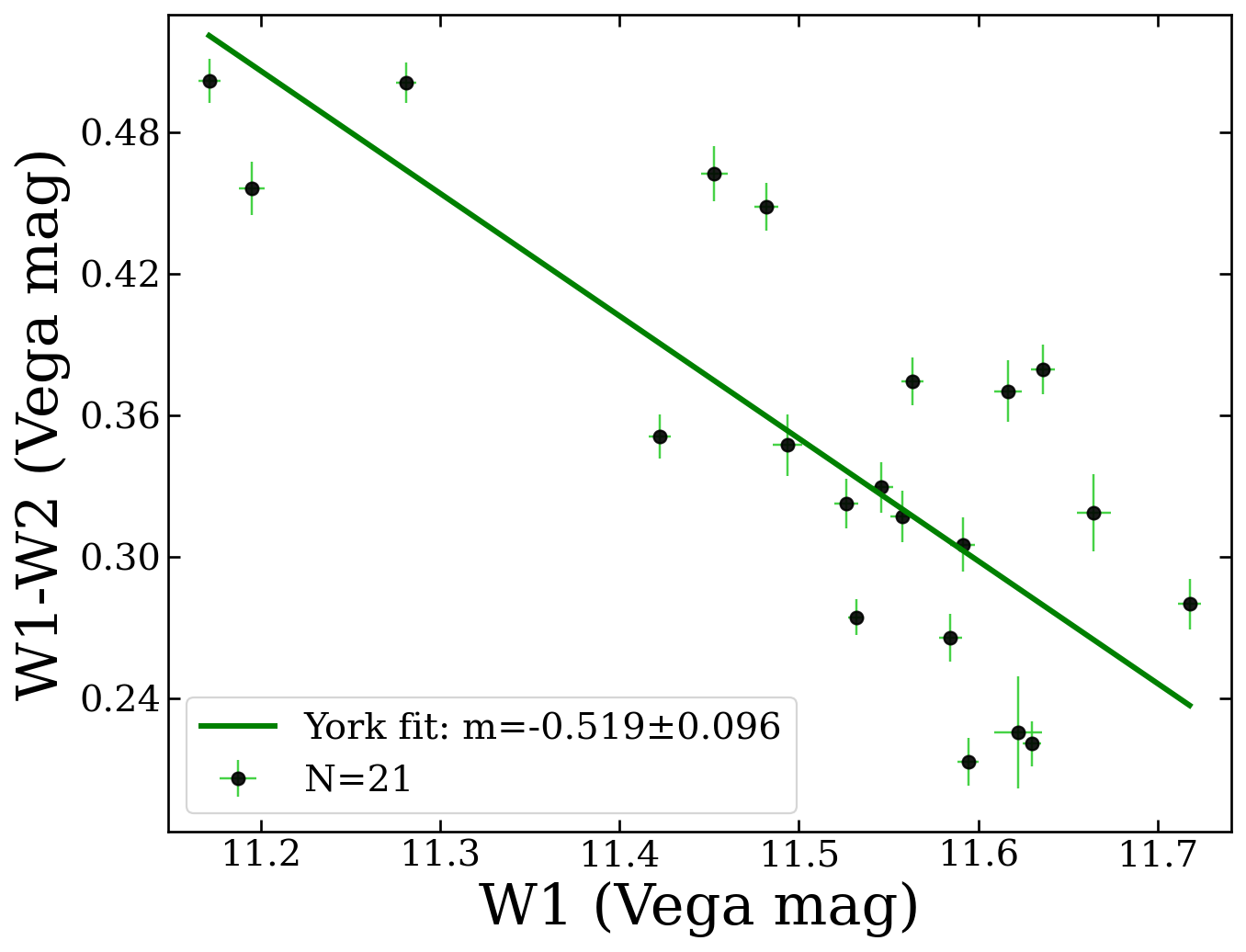}
\includegraphics[width=0.48\textwidth]
{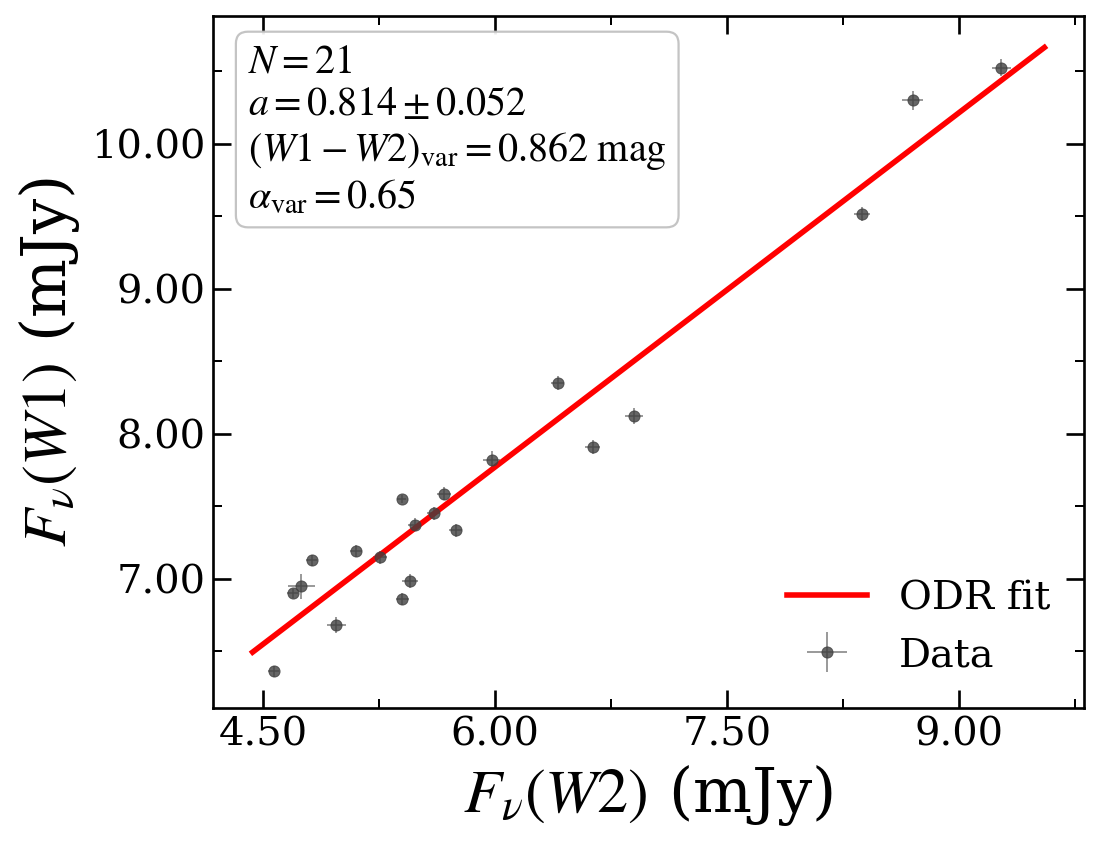}
\caption{Chromatic variability of IC~5287. Top: optical $g-r$ colour--magnitude and flux--flux relations. Bottom: corresponding WISE $W1-W2$ relations. The optical variability follows a BWB trend and the MIR variability a RWB trend. }
\label{fig:ic5287_colours}
\end{figure*}

Mrk~845 shows the larger optical variability amplitude, with central 90 per cent magnitude ranges of approximately 0.66, 0.38, and 0.30~mag in $g$, $r$, and $i$, respectively. Its $g-r$ relation has $m_{g-r}=0.479\pm0.009$
($r=0.935$, $p<10^{-6}$); restricting the sample to seeing $\leq2\arcsec$ gives a consistent slope of $0.481\pm0.014$. The 22 epoch-averaged WISE measurements give ${\rm d}(W1-W2)/{\rm d}W1=-0.263\pm0.039$ ($r=-0.799$, $p=8.2\times10^{-6}$). The corresponding spectral-index brightness dependences are ${\rm d}\alpha/{\rm d}(-g)=-1.503\pm0.029$ and ${\rm d}\alpha/{\rm d}(-W1)=+0.765\pm0.115$.

The principal flux--flux slopes are $a_{gr}=0.960\pm0.014$ and $a_{W1W2}=0.954\pm0.032$. The corresponding variable-component spectral indices are $\alpha_{\rm var}=0.14$ in $g-r$ and $0.15$ in $W1-W2$. In the optical, the correlated variable component is therefore substantially bluer than the median central-region emission, for which $\alpha_{\rm med}=2.173\pm0.015$. The complete chromatic and spectral-index diagnostics are summarised in Table~\ref{tab:chromatic_summary}. The principal optical and MIR colour--magnitude and flux--flux relations are shown in Fig.~\ref{fig:mrk845_colours}; additional seeing and colour tests are given in Appendix~\ref{app:mrk845_lag_tests}.

\begin{figure*}
\centering
\includegraphics[width=0.48\textwidth]
{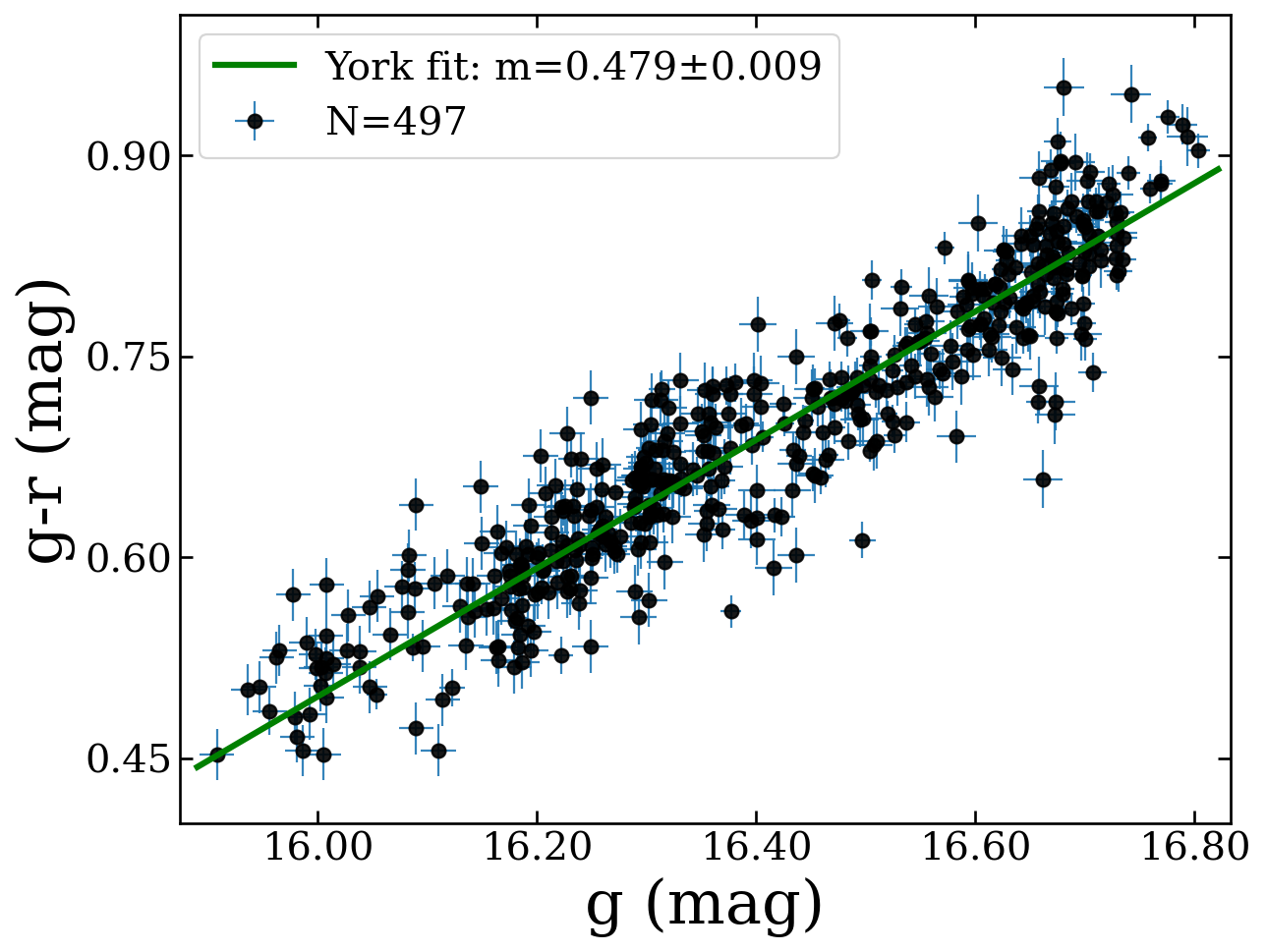}
\includegraphics[width=0.48\textwidth]
{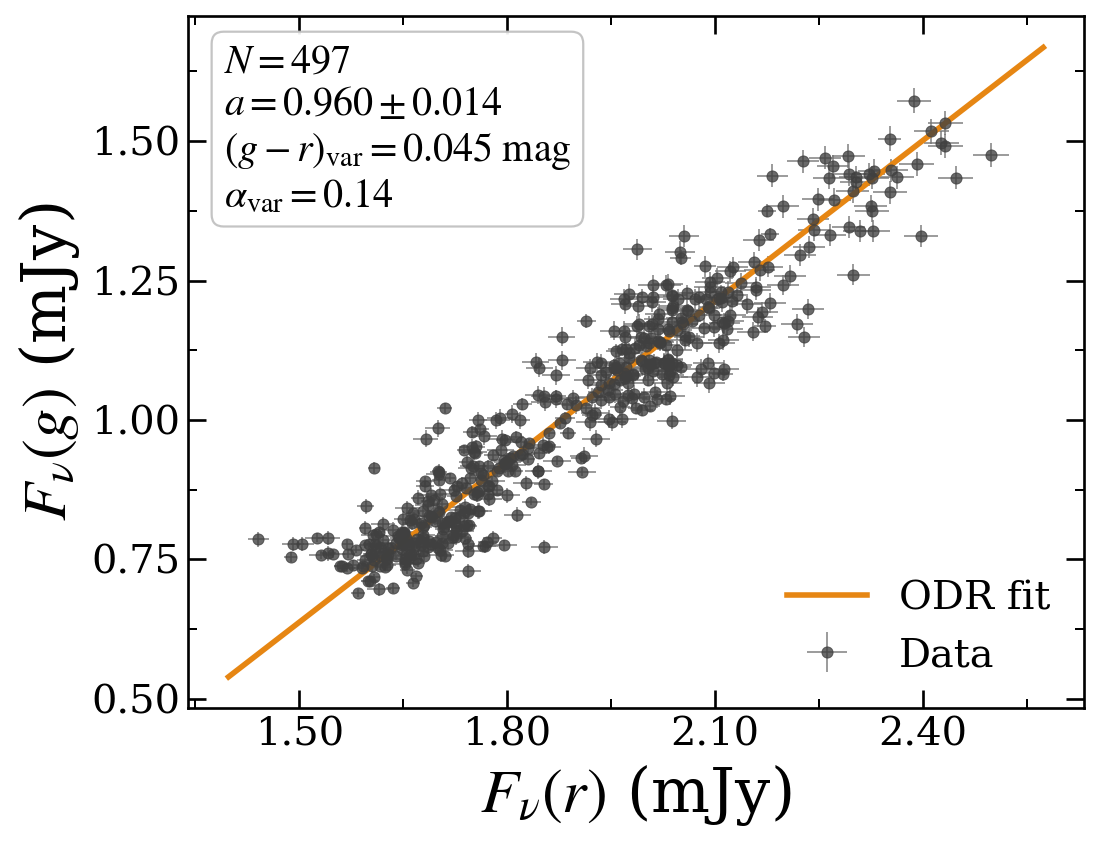}
\vspace{2mm}
\includegraphics[width=0.48\textwidth]
{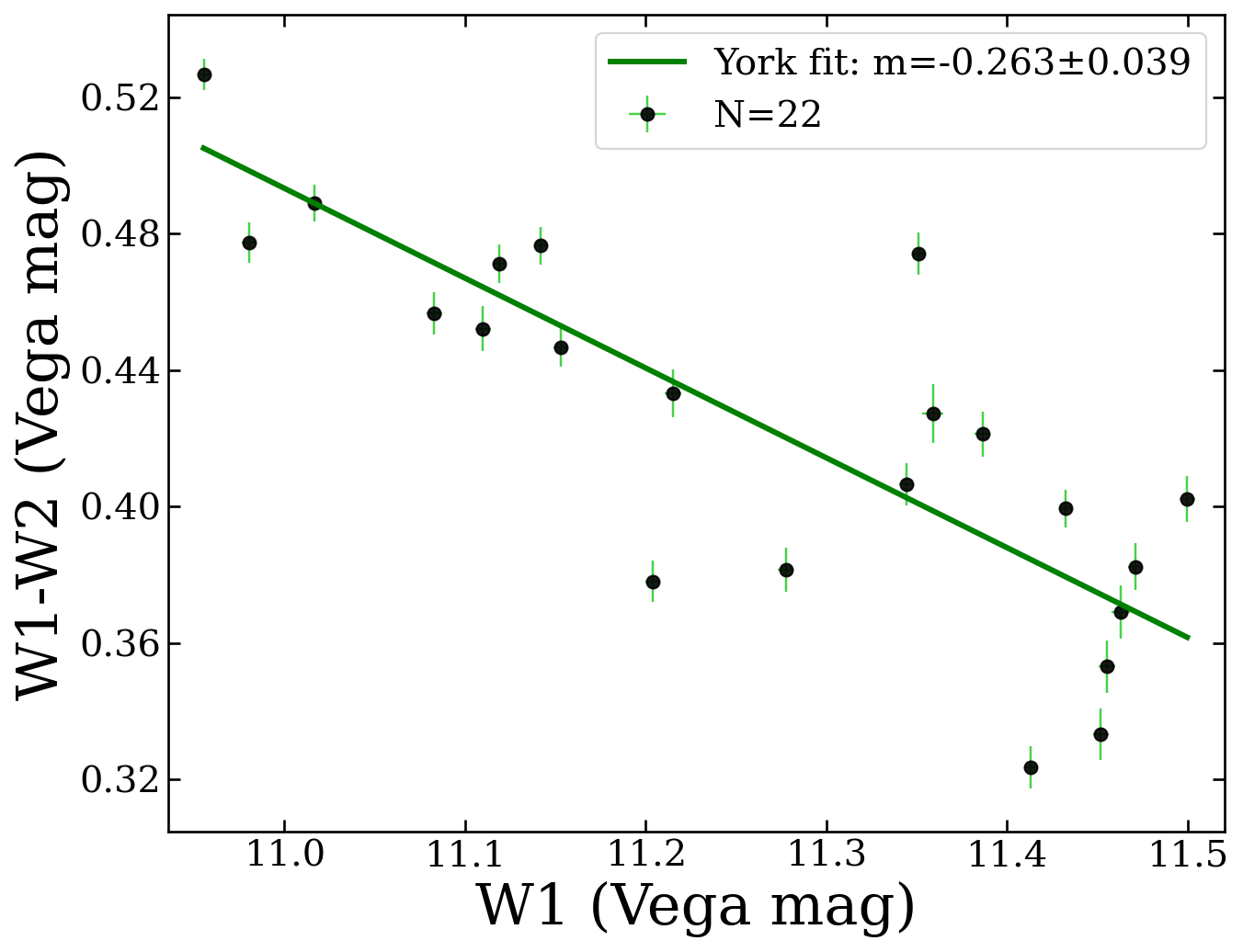}
\includegraphics[width=0.48\textwidth]
{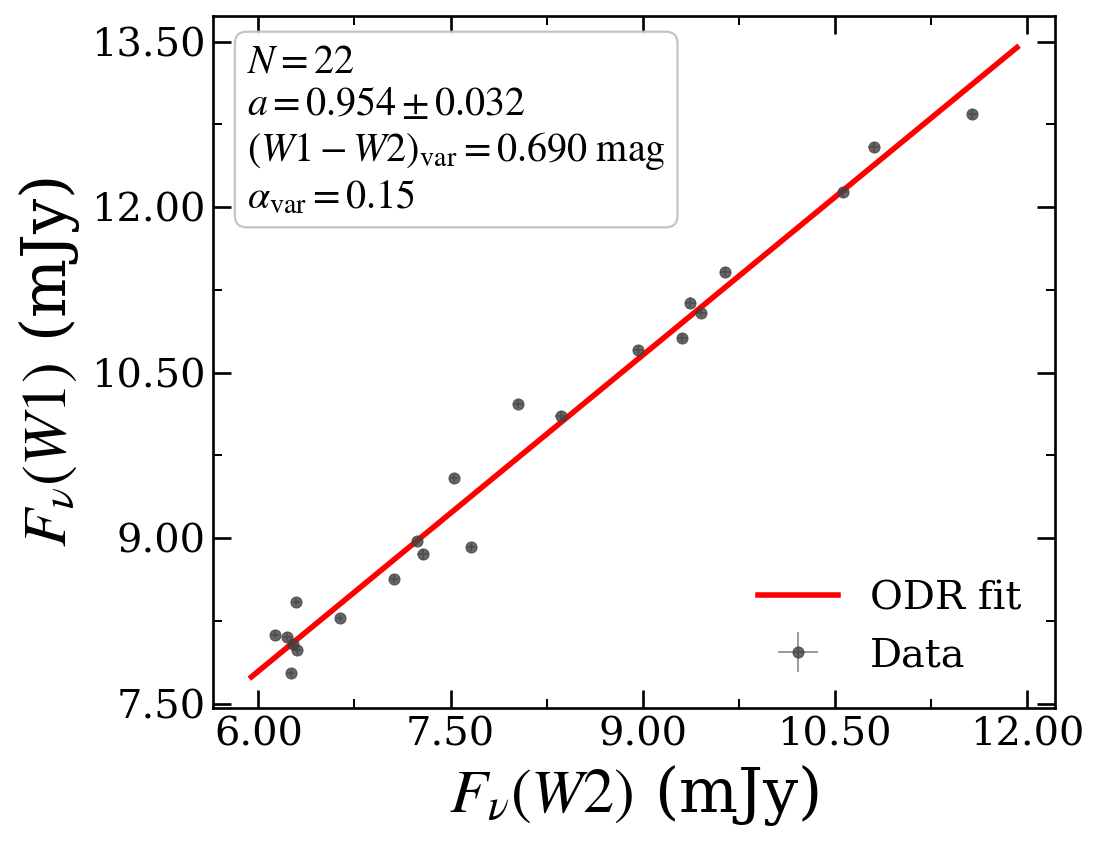}
\caption{Chromatic variability of Mrk~845. Top: optical $g-r$ colour--magnitude and flux--flux relations. Bottom: corresponding WISE $W1-W2$ relations. The source shows optical BWB and MIR RWB behaviour.}
\label{fig:mrk845_colours}
\end{figure*}

The chromatic diagnostics confirm the same qualitative combination of optical BWB and MIR RWB behaviour in both galaxies, with a substantially tighter optical colour--magnitude relation in Mrk~845. The colour-derived indices describe the observed central-region emission, while $\alpha_{\rm var}$ characterises the correlated variable component isolated by the flux--flux relation.

\begin{figure*}
\centering
\includegraphics[width=\textwidth]
{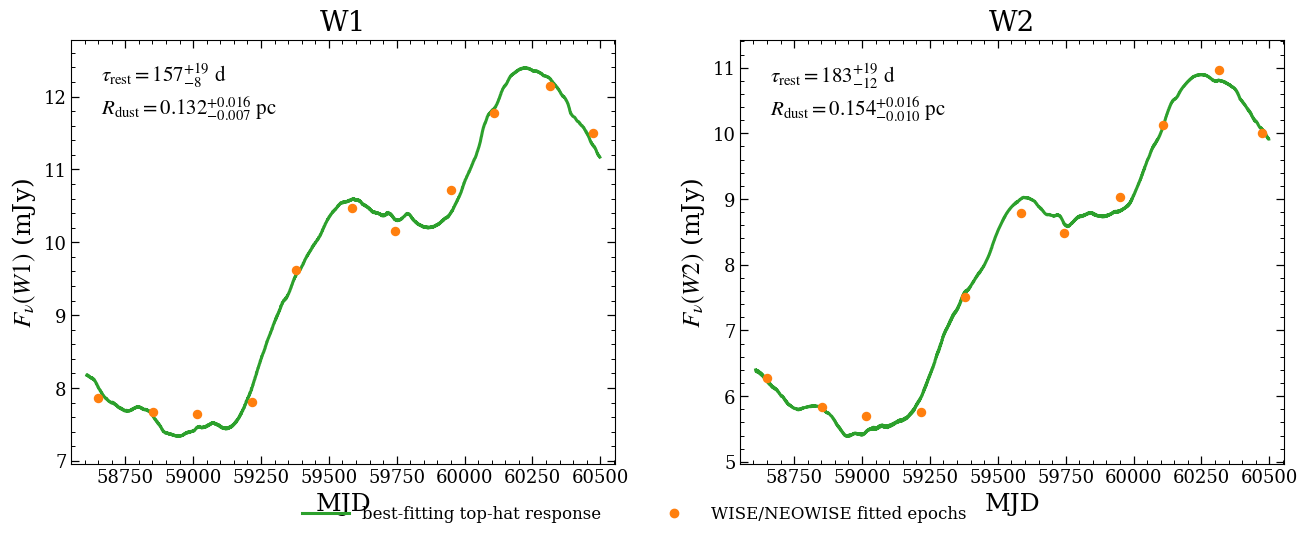}
\caption{Transfer-function reconstruction of the $W1$ and $W2$ variability of Mrk~845 using the ZTF $g$-band light curve as the optical driver. The curves show the MIR light curves reconstructed using the preferred top-hat transfer functions, while the points show the WISE/NEOWISE measurements included in the fit. Open early epochs mark measurements for which the complete positive-delay response is not covered by the optical driver. The adopted rest-frame delays are $157^{+19}_{-8}$~d in $W1$ and $183^{+19}_{-12}$~d in $W2$.}
\label{fig:mrk845_tf_reconstruction}
\end{figure*}

\subsubsection{Seasonal optical timescales in Mrk~845}
\label{subsec:mrk845_timescale_results}

The seasonal ZTF light curves of Mrk~845 show recurrent variability structure on timescales of several tens of days. The dominant seasonal GLS peak timescales span approximately 47--89~d in $g$ and 26--75~d in $r$.

Under the DRW Monte Carlo null model, the recurrence statistic gives $p_{\rm global}=6.0\times10^{-4}$ in $g$ and $p_{\rm global}=1.4\times10^{-2}$ in $r$. The conservative joint $g+r$ statistic gives $p_{\rm global}=8.7\times10^{-4}$. Leave-one-season-out tests show that the recurrent structure is not driven by a single observing season, while Gaussian-process model comparisons favour a stochastic DRW description over the tested DRW+SHO alternative.

We therefore interpret this behaviour as recurrent month-scale structure within stochastic optical variability, with no evidence for a stable periodicity. The complete seasonal GLS results, representative detrended light curves, Monte Carlo recurrence tests, leave-one-season-out tests, and stochastic-model comparisons are presented in Appendix~\ref{app:mrk845_timescale_tests}.

\subsection{Optical--MIR reverberation and hot-dust scales}
\label{subsec:results_reverberation}
 
Both galaxies show positive optical--MIR reverberation responses. We report the primary ICCF/FR--RSS results for IC~5287 and the transfer-function results for Mrk~845, following the estimator choices described in Section~\ref{subsec:lag_methods}.

For IC~5287, both MIR bands show broad positive responses on timescales
of order $10^2$~d. The 5000-realisation FR/RSS analysis gives
rest-frame centroid delays
\begin{equation}
\begin{aligned}
\tau_{\rm rest}(W1)
&=
106^{+24}_{-39}\ {\rm d},\\
\tau_{\rm rest}(W2)
&=
119^{+27}_{-25}\ {\rm d}.
\end{aligned}
\end{equation}
corresponding to characteristic dust-response radii
\begin{equation}
\begin{aligned}
R_{\rm dust}(W1)
&=
0.089^{+0.020}_{-0.033}\ {\rm pc},\\
R_{\rm dust}(W2)
&=
0.100^{+0.023}_{-0.021}\ {\rm pc}.
\end{aligned}
\end{equation}

The inferred delays remain stable across the tested accretion-disc-subtraction, ICCF, and lag-window configurations. Although the median $W2$ response occurs at a longer delay than the $W1$ response,
\begin{equation}
P[\tau(W2)>\tau(W1)]\simeq0.65,
\end{equation}
and the substantial overlap of the two lag distributions therefore provides only tentative evidence for wavelength ordering in IC~5287. The ICCF profiles, FR/RSS centroid distributions, and robustness tests are presented in Appendix~\ref{app:ic5287_phot_tests}.

For Mrk~845, the MIR variability was modelled as a delayed and temporally smoothed response to the ZTF $g$-band driver. Among the top-hat and Gaussian response functions tested, the top-hat model gives the lower BIC in both MIR bands. The corresponding best-fitting reconstruction is shown in Fig.~\ref{fig:mrk845_tf_reconstruction}.

After incorporating the accepted leave-one-epoch-out,
accretion-disc-subtraction, and interpolation-gap robustness tests,
we adopt the rest-frame delays
\begin{equation}
\begin{aligned}
\tau_{\rm rest}(W1)
&=
157^{+19}_{-8}\ {\rm d},\\
\tau_{\rm rest}(W2)
&=
183^{+19}_{-12}\ {\rm d}.
\end{aligned}
\end{equation}
corresponding to characteristic dust-response radii
\begin{equation}
\begin{aligned}
R_{\rm dust}(W1)
&=
0.132^{+0.016}_{-0.007}\ {\rm pc},\\
R_{\rm dust}(W2)
&=
0.154^{+0.016}_{-0.010}\ {\rm pc}.
\end{aligned}
\end{equation}

The $W2$ response occurs at a longer delay than the $W1$ response, with
\begin{equation}
P[\tau(W2)>\tau(W1)]=0.963.
\end{equation}
The median lag ratio is $\tau(W2)/\tau(W1)\simeq1.17$, and the wavelength ordering is retained across all paired robustness tests, supporting a wavelength-dependent MIR response structure in Mrk~845. The bootstrap distributions, response-function comparison, and additional robustness tests are presented in Appendix~\ref{app:mrk845_lag_tests}. The adopted delays, dust-response radii, and wavelength-ordering probabilities for both galaxies are summarised in Table~\ref{tab:reverb_summary}.

\begin{table*}
\centering
\caption{Principal optical--MIR reverberation measurements. The adopted estimator differs between the two sources because of their sampling and response morphology. All delays are given in the source rest frame.}
\label{tab:reverb_summary}
\renewcommand{\arraystretch}{1.15}
\begin{tabular*}{\textwidth}
{@{\extracolsep{\fill}}llccccc@{}}
\hline
Object &
Primary estimator &
$\tau_{W1}$ &
$\tau_{W2}$ &
$R_{W1}$ &
$R_{W2}$ &
$P[\tau(W2)>\tau(W1)]$ \\
&
&
(d) &
(d) &
(pc) &
(pc) &
\\
\hline
IC~5287 &
ICCF/FR--RSS &
$106^{+24}_{-39}$ &
$119^{+27}_{-25}$ &
$0.089^{+0.020}_{-0.033}$ &
$0.100^{+0.023}_{-0.021}$ &
$\simeq0.65$ \\
Mrk~845 &
transfer function &
$157^{+19}_{-8}$ &
$183^{+19}_{-12}$ &
$0.132^{+0.016}_{-0.007}$ &
$0.154^{+0.016}_{-0.010}$ &
$0.963$ \\
\hline
\end{tabular*}
\end{table*}
 
\subsection{Optical spectral properties}
\label{subsec:results_spectroscopy}
 
The SDSS spectral decompositions reveal different nuclear continuum contributions and broad-line reddening in the two galaxies. IC~5287 has the lower AGN continuum fraction at 5100~\AA\ and the larger broad-line Balmer decrement, whereas Mrk~845 has a larger AGN continuum contribution and a smaller inferred BLR colour excess.

For IC~5287, the AGN power-law contribution at 5100~\AA\ is
$f_{\rm AGN,5100}=0.143^{+0.008}_{-0.009}$,
and the fiducial host-subtracted nuclear luminosity is
\begin{equation}
\log
\left[
\frac{\lambda L_\lambda(5100)}
{{\rm erg\,s^{-1}}}
\right]
=
42.197^{+0.006}_{-0.010}.
\end{equation}
The decomposed AGN continuum has
\begin{equation}
F_\lambda\propto\lambda^\beta,
\qquad
\beta=+0.25.
\end{equation}
Additional stellar-continuum diagnostics, including the 4000-\AA\ break and host-fraction tests, are presented in Appendix~\ref{app:ic5287_line_diagnostics}.

The broad H$\beta$ profile of IC~5287 is strongly non-Gaussian. Because the conventional FWHM changes markedly depending on whether the inter-peak minimum lies above or below the half-maximum level, we adopt the line dispersion as the primary velocity measure:
\begin{equation}
\sigma_{\rm line}({\rm H}\beta)
=
2503^{+113}_{-98}\ {\rm km\,s^{-1}}.
\end{equation}
The sensitivity of the FWHM to the profile topology is illustrated in Fig.~\ref{fig:ic5287_hbeta_topology}; the full topology analysis and alternative width measurements are presented in Appendix~\ref{app:ic5287_hbeta}.

\begin{figure}
\centering
\includegraphics[width=\columnwidth]
{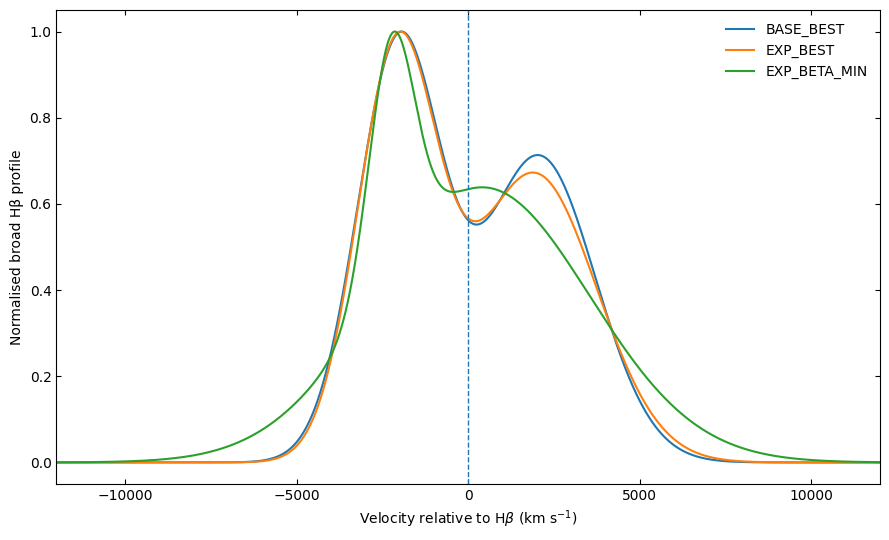}
\caption{Broad-H$\beta$ profile diagnostics for IC~5287. Representative accepted solutions illustrate the sensitivity of the conventional FWHM to whether the inter-peak minimum lies above or below the half-maximum level. Integrated profile measures such as $\sigma_{\rm line}$ and $W_{80}$ remain substantially more stable.}
\label{fig:ic5287_hbeta_topology}
\end{figure}

The broad Balmer decrement is
\begin{equation}
({\rm H}\alpha/{\rm H}\beta)_{\rm broad}
=
5.61^{+0.38}_{-0.35},
\end{equation}
giving
\begin{equation}
E(B-V)_{\rm BLR}
=
0.612^{+0.067}_{-0.066}\ {\rm mag}.
\end{equation}
The narrow-line decrement gives
\begin{equation}
E(B-V)_{\rm NLR}
=
0.378^{+0.104}_{-0.090}\ {\rm mag}.
\end{equation}

The decomposed narrow-line ratios place IC~5287 in the AGN region of the standard diagnostic diagrams. The adopted Seyfert-subtype criterion classifies 89.1 per cent of the Monte Carlo realisations as Sy~1.5 and 10.9 per cent as Sy~1.2. We therefore describe the 2000 SDSS spectrum as quantitatively more consistent with a Sy~1.5 state, while retaining the historical Sy~1.2 classification for the galaxy.

\begin{figure*}
\centering
\includegraphics[width=0.96\textwidth]
{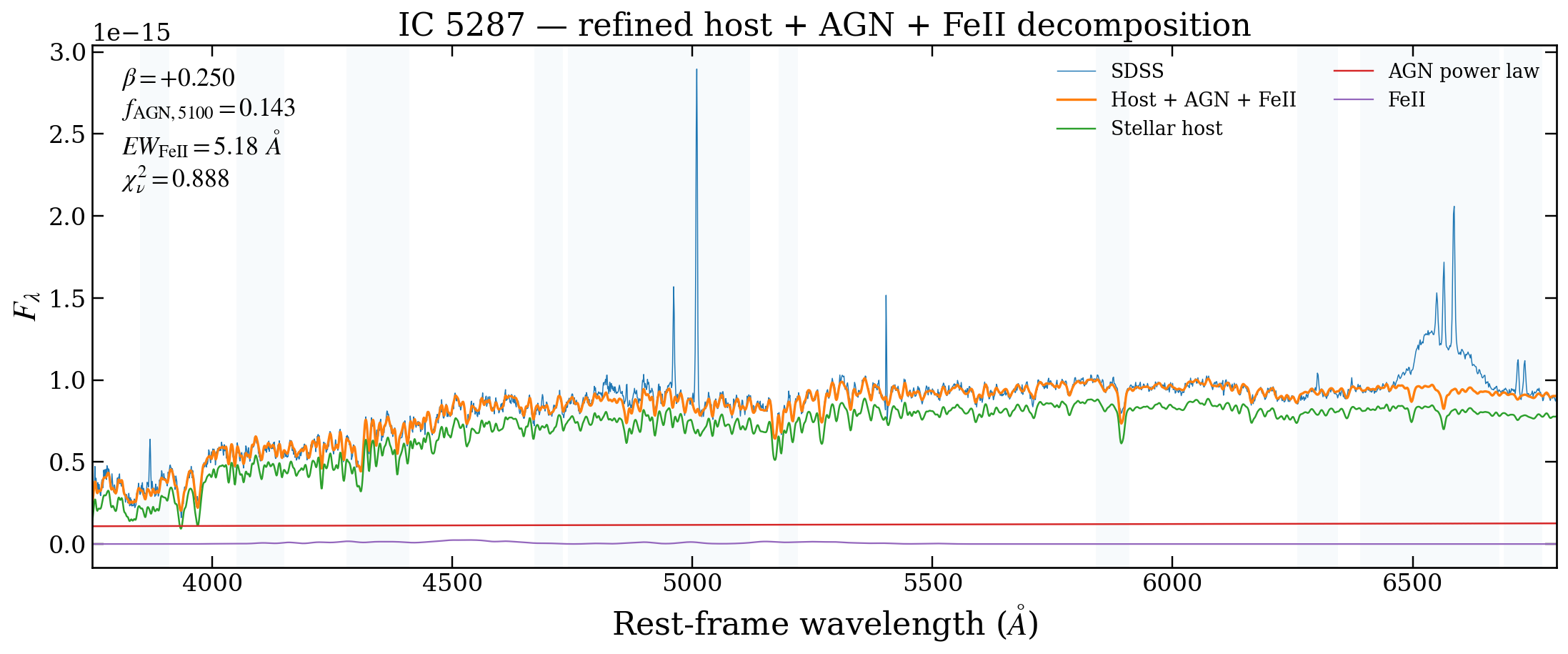}
\vspace{2mm}
\includegraphics[width=0.49\textwidth]
{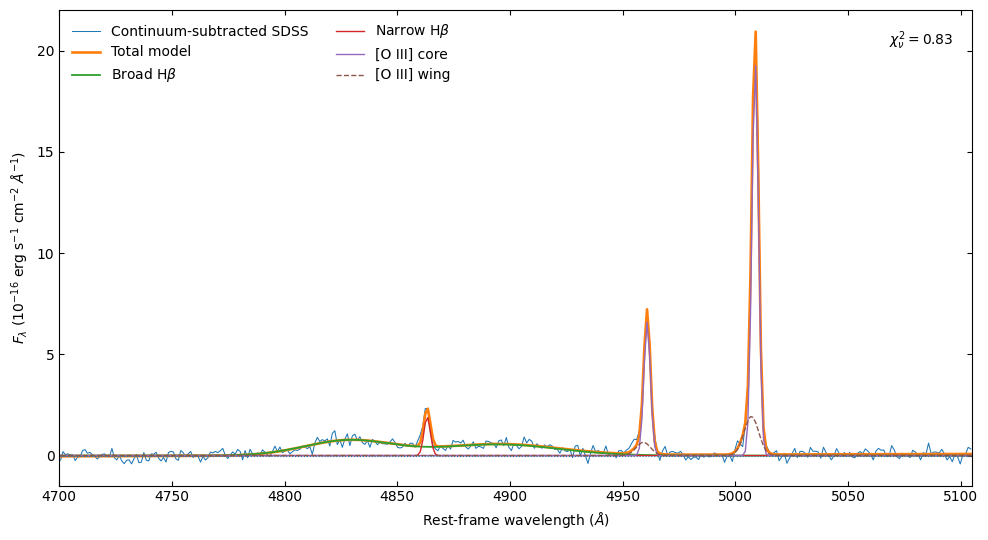}
\includegraphics[width=0.49\textwidth]
{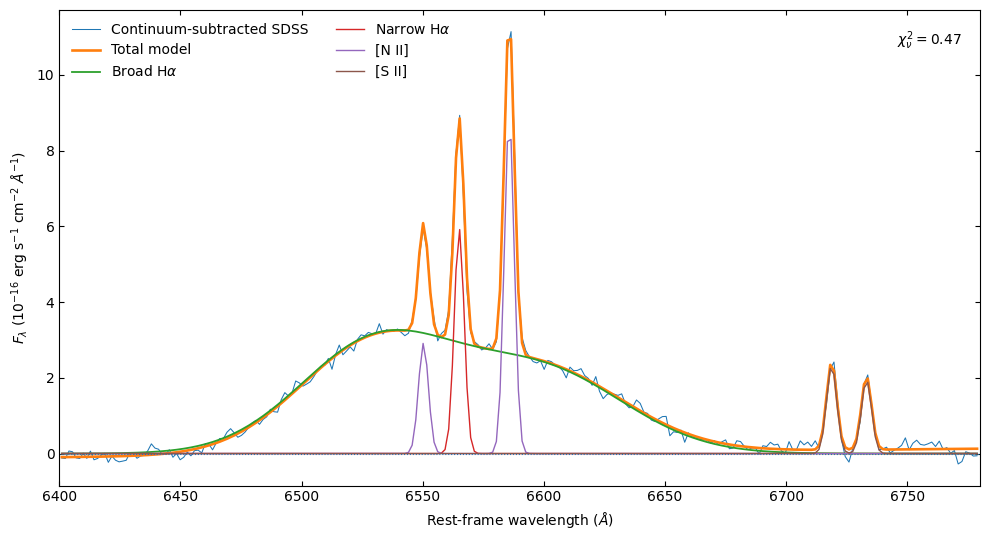}
\caption{Optical spectral decomposition of IC~5287. Top: Galactic-extinction-corrected SDSS spectrum with the stellar host, AGN power law, Fe~II, and total continuum components. Bottom left: continuum-subtracted H$\beta$+[O~III] region. Bottom right: H$\alpha$+[N~II]+[S~II] region.}
\label{fig:ic5287_spectrum}
\end{figure*}

The stellar-kinematic fit gives
\begin{equation}
\sigma_\star
=
128.5^{+5.3}_{-5.0}\ {\rm km\,s^{-1}}.
\end{equation}
Additional emission-line and stellar diagnostics are given in Appendix~\ref{app:ic5287_line_diagnostics}. The corresponding continuum and emission-line decomposition is shown in Fig.~\ref{fig:ic5287_spectrum}.

For Mrk~845, the continuum decomposition gives
\begin{equation}
f_{\rm AGN,5100}
=
0.401^{+0.020}_{-0.017},
\end{equation}
and
\begin{equation}
\log
\left[
\frac{\lambda L_\lambda(5100)}
{{\rm erg\,s^{-1}}}
\right]
=
42.958^{+0.025}_{-0.018}.
\end{equation}
The fiducial AGN power-law slope is
\begin{equation}
\beta
=
-0.050^{+0.069}_{-0.053}.
\end{equation}

The stellar velocity dispersion is
\begin{equation}
\sigma_\star
=
152.5^{+10.7}_{-14.8}\ {\rm km\,s^{-1}},
\end{equation}
consistent with the previously reported value of approximately
$139\pm17$~km~s$^{-1}$ \citep{Kompaniiets2025a}.

\begin{figure*}
\centering
\includegraphics[width=\textwidth]
{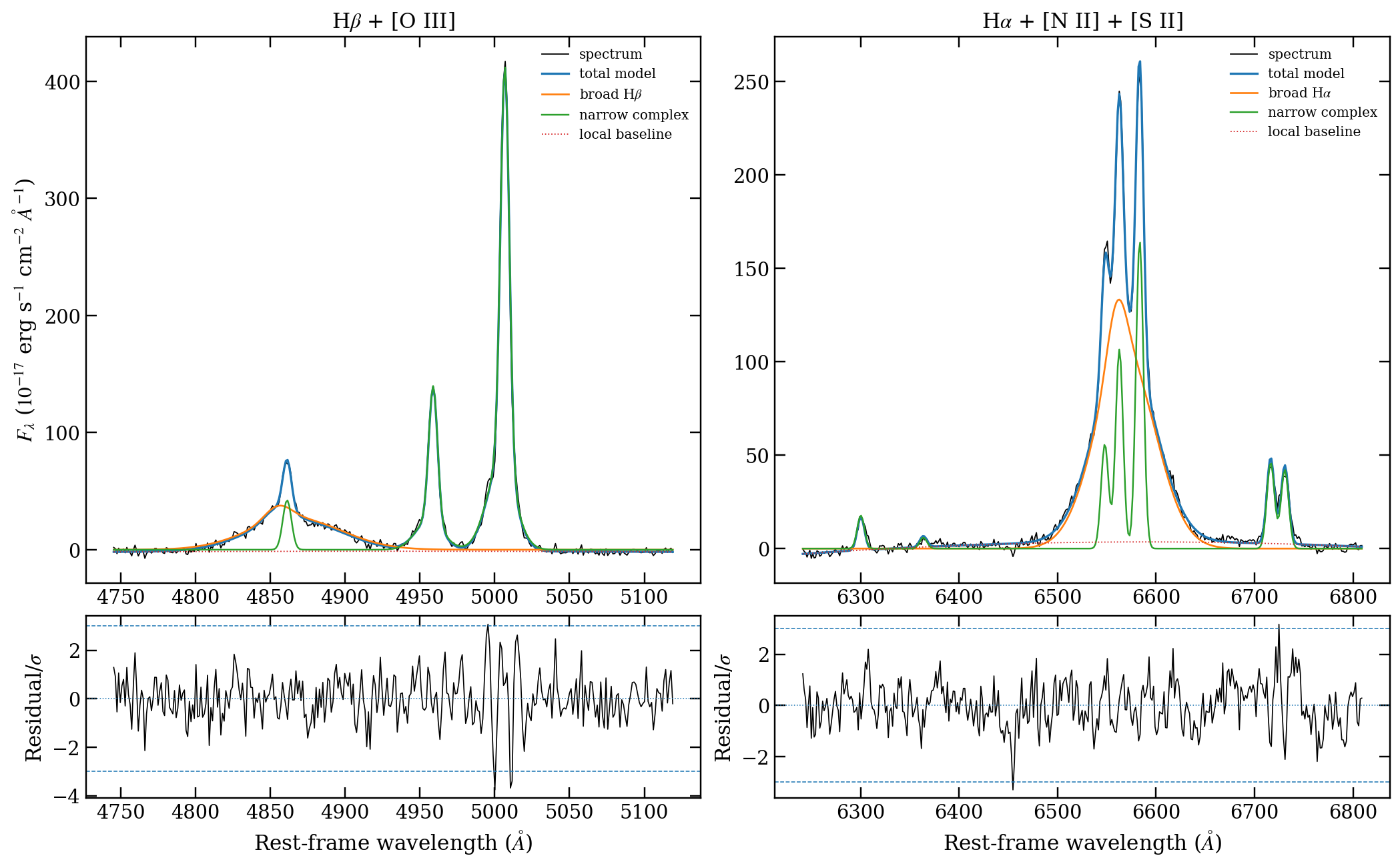}
\caption{Continuum-subtracted emission-line decomposition of the SDSS spectrum of Mrk~845. Left: H$\beta$+[O~III] region. Right: H$\alpha$+[N~II]+[S~II] region. The preferred model contains structured broad Balmer profiles and a blueshifted [O~III] wing. The lower panels show residuals in units of the spectral uncertainty.}
\label{fig:mrk845_linefit}
\end{figure*}

The preferred emission-line decomposition, shown in Fig.~\ref{fig:mrk845_linefit}, contains structured broad H$\beta$ and H$\alpha$ profiles together with an additional blueshifted [O~III] component. The broad-H$\beta$ line dispersion is
\begin{equation}
\sigma_{\rm line}({\rm H}\beta)
=
1814^{+56}_{-60}\ {\rm km\,s^{-1}}.
\end{equation}

The decomposed narrow-line ratios place Mrk~845 in the AGN/Seyfert region of all available BPT diagnostic diagrams, in contrast to the H~II-like classification obtained from the catalogue line measurements reported by \citet{Pulatova2015}. The \citet{Winkler1992} criterion classifies all Monte Carlo realisations as Sy~1.5. Detailed BPT distributions, broad-line model selection, and [O~III] kinematics are presented in Appendix~\ref{app:mrk845_spectral_tests}.

The broad Balmer decrement is
\begin{equation}
({\rm H}\alpha/{\rm H}\beta)_{\rm broad}
=
3.717^{+0.173}_{-0.142},
\end{equation}
corresponding to
\begin{equation}
E(B-V)_{\rm BLR}
=
0.197^{+0.046}_{-0.039}\ {\rm mag}.
\end{equation}
The narrow-line estimate,
\begin{equation}
E(B-V)_{\rm NLR}
=
0.060^{+0.092}_{-0.097},
\end{equation}
is consistent with zero.

The principal spectroscopic measurements for both galaxies are summarised in Table~\ref{tab:spectral_summary}.

\begin{table*}
\centering
\caption{
Principal spectroscopic measurements from the SDSS decompositions.
$L_{5100}$ denotes the fiducial host-subtracted AGN continuum luminosity
corrected for foreground Galactic extinction. Internal attenuation is
treated separately in the secondary reddening scenario. The activity
state refers specifically to the analysed SDSS epoch.
}
\label{tab:spectral_summary}

\renewcommand{\arraystretch}{1.15}
\begin{tabular*}{\textwidth}
{@{\extracolsep{\fill}}lcccccc@{}}
\hline
Object &
$f_{\rm AGN,5100}$ &
$\log L_{5100}$ &
$\sigma_\star$ &
$\sigma_{\rm H\beta}$ &
$E(B-V)_{\rm BLR}$ &
SDSS state \\
&
&
(erg s$^{-1}$) &
(km s$^{-1}$) &
(km s$^{-1}$) &
(mag) &
\\
\hline
IC~5287 &
$0.143^{+0.008}_{-0.009}$ &
$42.197^{+0.006}_{-0.010}$ &
$128.5^{+5.3}_{-5.0}$ &
$2503^{+113}_{-98}$ &
$0.612^{+0.067}_{-0.066}$ &
Sy~1.5 (89.1\%) \\
Mrk~845 &
$0.401^{+0.020}_{-0.017}$ &
$42.958^{+0.025}_{-0.018}$ &
$152.5^{+10.7}_{-14.8}$ &
$1814^{+56}_{-60}$ &
$0.197^{+0.046}_{-0.039}$ &
Sy~1.5 (100\%) \\
\hline
\end{tabular*}
\end{table*}

\subsection{Predicted BLR scales and the BLR--dust hierarchy}
\label{subsec:results_scales}
 
The host-subtracted continuum luminosities and broad-H$\beta$ line dispersions were used to predict the characteristic H$\beta$ BLR scales for both galaxies using the single-epoch spectroscopic calibration described in Section~\ref{subsec:blr_prediction_methods}. These predicted BLR scales are compared below with the independently measured MIR dust-reverberation radii.

For IC~5287, the fiducial host-subtracted continuum gives
\begin{equation}
\tau_{\rm BLR,pred}
=
4.75^{+3.62}_{-2.06}\ {\rm d}.
\end{equation}
Combining this value with the measured MIR radii gives
\begin{align}
\frac{R_{W1}}{R_{\rm BLR,pred}}
&=
20.28^{+18.46}_{-10.40},\\
\frac{R_{W2}}{R_{\rm BLR,pred}}
&=
24.75^{+20.61}_{-11.36}.
\end{align}

The elevated broad Balmer decrement motivates a secondary scenario in which the BLR colour excess is also applied to the optical AGN continuum. The resulting continuum slope is
\begin{equation}
\beta_{\rm int}
=
-2.37^{+0.28}_{-0.29},
\end{equation}
close to the standard long-wavelength thin-disc expectation
$F_\lambda\propto\lambda^{-7/3}$, while the corrected luminosity is
\begin{equation}
\log
\left[
\frac{\lambda L_\lambda(5100)}
{{\rm erg\,s^{-1}}}
\right]_{\rm int}
=
43.027^{+0.092}_{-0.089}.
\end{equation}

The corresponding predicted BLR lag becomes
\begin{equation}
\tau_{\rm BLR,pred,int}
=
11.68^{+8.52}_{-4.93}\ {\rm d},
\end{equation}
giving
\begin{align}
\frac{R_{W1}}{R_{\rm BLR,pred,int}}
&=
8.26^{+7.26}_{-4.17},\\
\frac{R_{W2}}{R_{\rm BLR,pred,int}}
&=
10.07^{+8.07}_{-4.52}.
\end{align}
The internal consistency of this attenuation scenario and the sensitivity to alternative broad-line velocity measures are examined in Appendix~\ref{app:ic5287_reddening}.

For Mrk~845, the fiducial host-subtracted continuum gives
\begin{equation}
\tau_{\rm BLR,pred}
=
8.12^{+5.67}_{-3.34}\ {\rm d}.
\end{equation}
Combining this predicted scale with the measured dust-reverberation
radii gives
\begin{align}
\frac{R_{\rm dust}(W1)}{R_{\rm BLR,pred}}
&=
19.7^{+14.0}_{-8.2},\\
\frac{R_{\rm dust}(W2)}{R_{\rm BLR,pred}}
&=
22.8^{+16.2}_{-9.4}.
\end{align}

Applying the same illustrative assumption,
$E(B-V)_{\rm cont}=E(B-V)_{\rm BLR}$, to Mrk~845 gives
\begin{equation}
\beta_{\rm int}
=
-0.754^{+0.169}_{-0.187},
\end{equation}
and
\begin{equation}
\log
\left[
\frac{\lambda L_\lambda(5100)}
{{\rm erg\,s^{-1}}}
\right]_{\rm int}
=
43.223^{+0.068}_{-0.060}.
\end{equation}

The corresponding predicted BLR lag increases to
\begin{equation}
\tau_{\rm BLR,pred,int}
=
10.85^{+7.50}_{-4.43}\ {\rm d},
\end{equation}
giving
\begin{align}
\frac{R_{\rm dust}(W1)}{R_{\rm BLR,pred,int}}
&=
14.8^{+10.4}_{-6.1},\\
\frac{R_{\rm dust}(W2)}{R_{\rm BLR,pred,int}}
&=
17.1^{+12.0}_{-7.0}.
\end{align}

The fiducial and attenuation-corrected BLR predictions and the corresponding dust-to-BLR radius ratios for both galaxies are summarised in Table~\ref{tab:radial_summary}.

\begin{table*}
\centering
\caption{Single-epoch spectroscopically predicted H$\beta$ BLR scales and the corresponding dust-to-predicted-BLR radius ratios. ``BLR-red.'' denotes the illustrative secondary scenario in which the broad-line colour excess is also applied to the AGN continuum, $E(B-V)_{\rm cont}=E(B-V)_{\rm BLR}$. The MIR dust radii are independently measured from optical--MIR reverberation and are unchanged between the two continuum scenarios.}
\label{tab:radial_summary}

\renewcommand{\arraystretch}{1.15}
\begin{tabular*}{\textwidth}
{@{\extracolsep{\fill}}lcccc@{}}
\hline
Object / continuum scenario &
$\log L_{5100}$ &
$\tau_{\rm BLR,pred}$ &
$R_{W1}/R_{\rm BLR}$ &
$R_{W2}/R_{\rm BLR}$ \\
&
(erg s$^{-1}$) &
(d) &
&
\\
\hline
IC~5287, fiducial &
$42.197^{+0.006}_{-0.010}$ &
$4.75^{+3.62}_{-2.06}$ &
$20.28^{+18.46}_{-10.40}$ &
$24.75^{+20.61}_{-11.36}$ \\
IC~5287, BLR-red. &
$43.027^{+0.092}_{-0.089}$ &
$11.68^{+8.52}_{-4.93}$ &
$8.26^{+7.26}_{-4.17}$ &
$10.07^{+8.07}_{-4.52}$ \\
Mrk~845, fiducial &
$42.958^{+0.025}_{-0.018}$ &
$8.12^{+5.67}_{-3.34}$ &
$19.7^{+14.0}_{-8.2}$ &
$22.8^{+16.2}_{-9.4}$ \\
Mrk~845, BLR-red. &
$43.223^{+0.068}_{-0.060}$ &
$10.85^{+7.50}_{-4.43}$ &
$14.8^{+10.4}_{-6.1}$ &
$17.1^{+12.0}_{-7.0}$ \\[3pt]
\hline
\end{tabular*}
\end{table*}

The reported radius ratios and uncertainty intervals are obtained from the fully propagated distributions. Their median values therefore need not equal the ratios of the separately quoted median delays.

For comparison with the population
$R_{\rm dust}$--$L_{5100}$ relations of \citet{Ayubinia2025}, we apply
the wavelength--redshift correction defined in
equation~\ref{eq:dust_redshift_corr}. At the fiducial host-subtracted
luminosities, the resulting residuals from the free-slope relations are
\begin{equation}
\begin{aligned}
{\rm IC~5287:}\qquad
\Delta\log R_{\rm dust}(W1)
&\simeq +0.22, &
\Delta\log R_{\rm dust}(W2)
&\simeq +0.09,\\
{\rm Mrk~845:}\qquad
\Delta\log R_{\rm dust}(W1)
&\simeq +0.13, &
\Delta\log R_{\rm dust}(W2)
&\simeq +0.03.
\end{aligned}
\end{equation}
Both objects therefore remain within the intrinsic scatter of the
corresponding dust size--luminosity relations.

\begin{figure}
\centering
\includegraphics[width=\columnwidth]
{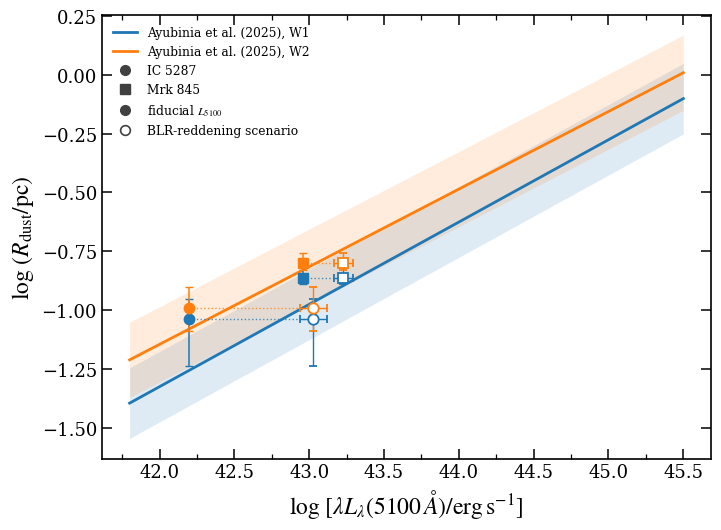}
\caption{
Hot-dust response radii of IC~5287 and Mrk~845 compared with the
$W1$ and $W2$ dust size--luminosity relations of
\citet{Ayubinia2025}. For this population comparison, the measured
optical--MIR delays are corrected for both cosmological time dilation
and the wavelength--redshift dependence using
$\tau_{\rm corr}=\tau_{\rm obs}(1+z)^{-0.26}$, following
\citet{Ayubinia2025}. Filled symbols show the fiducial
host-subtracted 5100-\AA\ luminosities, while open symbols show the
illustrative BLR-reddening continuum scenarios. Dotted lines connect
the two luminosity estimates for the same measured dust response. The
dust reverberation measurement itself is unchanged between the two
continuum scenarios.
}
\label{fig:current_rdust_l5100}
\end{figure}

The implications of the absolute dust scales, internal attenuation, and the resulting BLR--dust radial hierarchies are discussed in Section~\ref{sec:discussion}.
 
\section{Discussion}
\label{sec:discussion}

When placed on the population $R_{\rm dust}$--$L_{5100}$ relations
using the wavelength--redshift correction adopted for that comparison,
the dust-response scales of IC~5287 and Mrk~845 are consistent with
the broader type-1 AGN population once the measurement uncertainties
and intrinsic scatter of the relations are taken into account. The
interpretation of the comparatively large dust-to-BLR radius ratios
is particularly sensitive to the single-epoch spectroscopic prediction
of the BLR scale and to the quantities entering that prediction, most
notably the nuclear continuum luminosity. This sensitivity is strongest
for IC~5287, whose optical spectrum is strongly host dominated and
shows substantial broad-line reddening.

\subsection{Hot-dust reverberation and wavelength-dependent structure}
\label{subsec:discussion_dust}

IC~5287 shows a characteristic optical--MIR dust response on a scale of approximately 0.1~pc. The median $W2$ delay is longer than the $W1$ delay, as expected qualitatively if the longer-wavelength emission preferentially samples cooler dust at larger characteristic radii. However, the substantial overlap between the two lag distributions, with $P[\tau(W2)>\tau(W1)]\simeq0.65$, means that the wavelength ordering is only tentative for this object. Overlapping wavelength-dependent lag distributions have also been reported in individual dust-reverberation campaigns \citep[e.g.][]{Mandal2021}.

The wavelength dependence is more strongly constrained in Mrk~845. Its $W2$ response occurs at a longer delay than the $W1$ response, with $P[\tau(W2)>\tau(W1)]=0.963$, and the ordering is retained across the paired robustness tests. The median lag ratio, $\tau(W2)/\tau(W1)\simeq1.17$, is close to the population value implied by the characteristic $R_K:R_{W1}:R_{W2}\simeq1.0:1.5:1.8$ relation reported by \citet{Mandal2024}, corresponding to $R_{W2}/R_{W1}\simeq1.2$. Thus, the wavelength dependence inferred for Mrk~845 is consistent with the radial stratification expected for a thermally reprocessing dust distribution.

For Mrk~845, the transfer-function modelling reproduces the principal
long-term MIR variations and yields characteristic delays that remain
stable across the accepted robustness tests. The fitted response widths
are considerably less stable and frequently approach the allowed fitting
boundary. We therefore interpret the measured delays as response-weighted
characteristic dust scales and treat the fitted widths solely as
phenomenological descriptors of the temporal response.

The measured dust-response scales are consistent with the population
$R_{\rm dust}$--$L_{5100}$ relations of \citet{Ayubinia2025} once the
measurement uncertainties and intrinsic scatter are taken into account.
Their comparatively large $R_{\rm dust}/R_{\rm BLR,pred}$ ratios
therefore primarily reflect the spectroscopically predicted BLR scales
and the quantities entering those predictions, while the measured
dust-reverberation radii are consistent with those of the broader AGN
population.
This distinction is important because $R_{\rm dust}$ is measured
directly from the optical--MIR reverberation response, whereas
$R_{\rm BLR,pred}$ is inferred from single-epoch spectroscopy. The
resulting BLR--dust hierarchy is consequently more sensitive to the
continuum luminosity and broad-line kinematics entering the BLR
prediction.

The SDSS spectroscopy and the ZTF/WISE reverberation measurements
sample widely separated source epochs. The measured $R_{\rm dust}$
values and the single-epoch quantities entering $R_{\rm BLR,pred}$
therefore characterise different nuclear states. This temporal
separation is relevant because the dust-response scale can retain a
dependence on the preceding luminosity history
\citep{Koshida2014}. Changes in the inner dust scale following
substantial luminosity variations have been observed in other AGNs,
including Mrk~590 \citep{Kokubo2020}. Source-state evolution may thus
contribute to the positions of IC~5287 and Mrk~845 in the
$R_{\rm dust}$--$L_{5100}$ plane and to their inferred
dust-to-predicted-BLR radius ratios.

\subsection{Internal attenuation and the BLR--dust hierarchy}
\label{subsec:discussion_extinction}

The influence of the optical continuum estimate is most evident in IC~5287. Its SDSS spectrum is strongly host dominated, with only about 14 per cent of the decomposed continuum at 5100~\AA\ attributed to the AGN, while the broad Balmer decrement implies substantial BLR reddening,
\begin{equation}
E(B-V)_{\rm BLR}
=
0.612^{+0.067}_{-0.066}\ {\rm mag}.
\end{equation}
The host-subtracted SDSS continuum is red, and the independently determined flux--flux relation shows that the correlated optical variable component is also red in the observed frame despite the overall BWB behaviour. Taken together, these diagnostics support the possibility that nuclear attenuation contributes in addition to the strong host dilution, although they do not independently determine the reddening of the optical continuum source.

For the fiducial host-subtracted continuum, the single-epoch prediction gives an H$\beta$ BLR scale that is small relative to the measured MIR reverberation radii, resulting in $R_{\rm dust}/R_{\rm BLR,pred}$ ratios of approximately 20--25. These values are larger than the characteristic hierarchy found in AGNs with directly measured H$\beta$ reverberation lags; for example, \citet{Chen2023} obtained
\begin{equation}
R_{\rm BLR}:R_K:R_{W1}:R_{W2}\simeq1:6.2:9.2:11.2.
\end{equation}
This comparison is not strictly like-for-like, however, because the BLR scale in that population hierarchy is reverberation measured, whereas $R_{\rm BLR,pred}$ in IC~5287 is inferred from single-epoch spectroscopy. The latter depends explicitly on the adopted nuclear continuum luminosity and broad-H$\beta$ velocity scale.

To quantify the sensitivity of the inferred hierarchy to this effect, we considered the illustrative scenario
\begin{equation}
E(B-V)_{\rm cont}
=
E(B-V)_{\rm BLR}.
\end{equation}
For IC~5287, this assumption changes the inferred continuum slope to
$\beta_{\rm int}\simeq-2.37$, close to the standard long-wavelength
thin-disc expectation $F_\lambda\propto\lambda^{-7/3}$
\citep{ShakuraSunyaev1973}, and increases the inferred intrinsic
$L_{5100}$ by approximately 0.83~dex. The resulting predicted H$\beta$ lag increases from approximately 4.8 to 11.7~d, reducing the dust-to-predicted-BLR radius ratios from approximately 20--25 to 8--10. The measured MIR reverberation radii themselves are unaffected by this continuum-reddening assumption.

We treat this attenuation prescription as an illustrative scenario for
the continuum reddening. In a structured circumnuclear medium, the
optical continuum source and the BLR can probe different effective
sightlines. The archival Chandra spectrum of IC~5287 gives an absorbing
column of $N_{\rm H}\simeq3.9\times10^{21}$~cm$^{-2}$
\citep{Kompaniiets2025a}, providing independent evidence for absorbing
material along the nuclear line of sight. The X-ray absorbing column
and the optical BLR reddening may trace distinct components with
different gas-to-dust properties.

Mrk~845 provides a useful contrast. Its broad-line reddening is more modest,
\begin{equation}
E(B-V)_{\rm BLR}
=
0.197^{+0.046}_{-0.039}\ {\rm mag},
\end{equation}
while the narrow-line reddening is consistent with zero. Applying the same illustrative assumption, $E(B-V)_{\rm cont}=E(B-V)_{\rm BLR}$, increases the predicted H$\beta$ lag only from approximately 8.1 to 10.9~d and reduces the dust-to-predicted-BLR radius ratios from approximately 20--23 to 15--17. The inferred attenuation-corrected continuum remains substantially redder than the standard long-wavelength thin-disc expectation. Thus, the same reddening prescription has a considerably smaller effect on the inferred BLR--dust hierarchy of Mrk~845 than on that of IC~5287.

The contrasting behaviour of the two galaxies illustrates how an apparently unusual dust-to-BLR hierarchy can arise without an unusually large measured dust radius. In a comparison based on single-epoch spectroscopy, host dilution and internal attenuation modify the nuclear continuum luminosity and hence the predicted BLR scale, while the independently measured dust-reverberation radius remains unchanged. The inferred $R_{\rm dust}/R_{\rm BLR,pred}$ hierarchy can therefore be highly sensitive to the treatment of the optical nuclear continuum, particularly in strongly host-dominated and reddened systems such as IC~5287.

\subsection{Optical spectral state and variability}
\label{subsec:discussion_spectra}

The SDSS spectra place both nuclei in intermediate type-1 states at their respective observing epochs. For IC~5287, the 2000 SDSS spectrum is quantitatively more consistent with Sy~1.5, whereas the galaxy has historically been classified as Sy~1.2 \citep{Pietsch1998}. This difference should not by itself be interpreted as evidence for a discrete spectral-type transition, because the available spectra are non-contemporaneous and differ in aperture, calibration, and analysis procedure. The smaller broad Balmer decrement reported by \citet{Kollatschny2008} likewise indicates that homogeneous multi-epoch spectroscopy would be required to establish whether the broad-line spectrum and its reddening vary systematically with nuclear state.

Mrk~845 is likewise classified as Sy~1.5 at the SDSS epoch. Its decomposed narrow-line ratios occupy the AGN/Seyfert regions of the available BPT diagrams, in contrast to the H~II-like catalogue classification reported by \citet{Pulatova2015}. This difference illustrates the importance of separating the nuclear emission-line components from the underlying continuum and blended line structure when assigning diagnostic classifications in composite host--AGN spectra. The preferred decomposition also requires a blueshifted asymmetric [O~III] wing, indicating a kinematically disturbed ionised-gas component.

Both galaxies become optically bluer as they brighten and show redder-when-brighter behaviour in the MIR, but the spectra of their correlated optical variable components differ substantially. IC~5287 retains a red observed variable-component spectrum despite its relative hardening with increasing brightness, whereas Mrk~845 shows a substantially bluer variable component. For IC~5287, the red variable-component spectrum is also qualitatively consistent with the attenuation-sensitive picture discussed in Section~\ref{subsec:discussion_extinction}, although the photometric colours alone do not determine the continuum reddening. Together with the positive optical--MIR reverberation delays, the chromatic variability is consistent with a variable optical continuum whose variations are subsequently reprocessed by circumnuclear dust.

Mrk~845 additionally shows recurrent optical structure on timescales
of several tens of days. The DRW-based Monte Carlo analysis detects
significant recurrence, and the leave-one-season-out tests show that
the signal persists across the observing seasons. The DRW model also
has the lower BIC in all season--filter combinations considered in the
Gaussian-process analysis. We therefore interpret the signal as
recurrent month-scale stochastic optical variability. Evolving or
recurrent characteristic timescales have been reported in other AGNs,
including Zw~229--15 \citep{Phillipson2020}, while population studies
have linked stochastic optical variability timescales to accretion-disc
thermal physics and black-hole mass \citep{Burke2021}.

The dedicated IAC80 monitoring detected no statistically significant intraday optical variability in either nucleus during the sampled epochs. This null result is specific to the temporal coverage and sensitivity of the available observations and does not exclude intraday variability at other epochs or below the present detection threshold.

Taken together, the spectroscopy and variability diagnostics show that the two isolated Seyfert nuclei occupy broadly similar intermediate type-1 states but differ substantially in host dilution, reddening, optical variable-component colour, and short-to-intermediate-timescale behaviour. These differences provide the nuclear-state context for interpreting their spectroscopically predicted BLR scales and optical--MIR reverberation measurements.

\section{Conclusions}
\label{sec:conclusions}

We have combined long-baseline optical and MIR variability with host-decomposed single-epoch spectroscopy to investigate the hot-dust reverberation scales and the inferred BLR--dust radial hierarchy in the isolated Seyfert galaxies IC~5287 and Mrk~845. ZTF and WISE/NEOWISE monitoring provides direct optical--MIR reverberation measurements of the hot-dust response, while the archival SDSS spectra provide the nuclear continuum and broad-H$\beta$ quantities used to predict the BLR scale. Dedicated IAC80 monitoring provides an additional test for intraday optical variability.

Both nuclei show delayed MIR responses consistent with thermal
reprocessing by circumnuclear dust. For IC~5287, the rest-frame delays
are $106^{+24}_{-39}$~d in $W1$ and $119^{+27}_{-25}$~d in $W2$,
corresponding to characteristic dust-response radii of
$0.089^{+0.020}_{-0.033}$ and $0.100^{+0.023}_{-0.021}$~pc,
respectively. The median $W2$ delay is longer, but the wavelength
ordering remains tentative, with
$P[\tau(W2)>\tau(W1)]\simeq0.65$. For Mrk~845, the corresponding
delays are $157^{+19}_{-8}$ and $183^{+19}_{-12}$~d, giving
dust-response radii of $0.132^{+0.016}_{-0.007}$ and
$0.154^{+0.016}_{-0.010}$~pc. In this object the longer $W2$ response
is more strongly supported, with
$P[\tau(W2)>\tau(W1)]=0.963$. When compared with the WISE dust
size--luminosity relations using the corresponding
wavelength--redshift correction, the dust-response scales of both
galaxies are consistent with the population relations once the
measurement uncertainties and intrinsic scatter are taken into
account.

The spectroscopic comparison shows that the comparatively large inferred
BLR--dust hierarchies are primarily set by the predicted BLR scales.
Using the fiducial host-subtracted continua, the single-epoch
spectroscopic predictions give
$\tau_{\rm BLR,pred}\simeq4.8$~d for IC~5287 and
$\simeq8.1$~d for Mrk~845, yielding
$R_{\rm dust}/R_{\rm BLR,pred}$ ratios of approximately 20--25 and
20--23, respectively. The measured hot-dust reverberation radii
remain consistent with the population dust size--luminosity relations,
while the relatively small predicted BLR scales produce the large
radius ratios.

IC~5287 is strongly host dominated and shows substantial broad-line
reddening,
$E(B-V)_{\rm BLR}=0.612^{+0.067}_{-0.066}$~mag. Under the illustrative
assumption $E(B-V)_{\rm cont}=E(B-V)_{\rm BLR}$, the inferred intrinsic
continuum becomes substantially bluer, the predicted H$\beta$ BLR lag
increases from approximately 4.8 to 11.7~d, and the
dust-to-predicted-BLR radius ratios decrease from approximately
20--25 to 8--10. Mrk~845 shows more moderate broad-line reddening,
$E(B-V)_{\rm BLR}=0.197^{+0.046}_{-0.039}$~mag; applying the same
assumption changes its predicted BLR lag only from approximately
8.1 to 10.9~d and reduces the corresponding radius ratios from
approximately 20--23 to 15--17. The contrasting response of the two
objects demonstrates the sensitivity of single-epoch BLR predictions
to the treatment of the nuclear continuum, particularly in strongly
host-dominated and reddened nuclei.

At their respective SDSS epochs, both nuclei are quantitatively
consistent with intermediate Sy~1.5 states. Both also exhibit BWB
optical variability and RWB MIR behaviour. Their correlated optical
components differ substantially: IC~5287 retains a red observed
variable-component spectrum, while Mrk~845 shows a much bluer variable
component. In Mrk~845, the seasonal analysis identifies recurrent
structured stochastic variability on timescales of several tens of
days. The IAC80 observations show no statistically significant
intraday optical variability in either galaxy during the sampled
epochs.

Overall, these results emphasise the need to distinguish directly measured hot-dust reverberation scales from BLR radii predicted from single-epoch spectroscopy when constructing BLR--dust radial hierarchies. The latter remain sensitive to host dilution, internal attenuation, broad-line kinematics, and the nuclear continuum luminosity entering the adopted BLR calibration. Contemporaneous optical spectroscopy and reverberation monitoring, together with denser infrared sampling, will be required to reduce these systematic uncertainties and to determine whether the large dust-to-predicted-BLR ratios found here persist when both radial scales are measured for the same nuclear state. The present results show that an apparently extended BLR--dust hierarchy need not imply an anomalously large hot-dust radius.

\section*{Acknowledgements}
This study was supported by the Research programme for young scientists of the National Academy of Sciences of Ukraine for 2025--2026 (Project ID 0125U002943). We used observational data obtained with the IAC80 telescope, operated by the Instituto de Astrofísica de Canarias at the Observatorio del Teide, Tenerife, Spain. We thank the support astronomers and night assistants for their assistance with the observations. This work makes use of publicly available data from SDSS, ZTF, \textit{WISE}/\textit{NEOWISE}, and the NASA/IPAC Infrared Science Archive (IRSA).

\section*{Data availability}
The ZTF photometric data used in this study are publicly available through the Zwicky Transient Facility archive. The WISE and NEOWISE photometric data are publicly available through the NASA/IPAC Infrared Science Archive (IRSA), and the archival optical spectra are publicly available through the Sloan Digital Sky Survey (SDSS).

The dedicated IAC80/CAMELOT2 observations were obtained under observing programmes awarded to the authors and were carried out in service mode with the support of the Instituto de Astrofísica de Canarias observing staff. The reduced IAC80 data products and the derived measurements produced in this work are available from the corresponding author upon reasonable request.

The comparison-star sequences used for the IAC80/CAMELOT2 differential photometry, including J2000 coordinates, APASS DR9 multiband catalogue photometry, and static finding charts, are publicly available in the \textit{AGN Reference Fields} dataset on Zenodo \citep{Izviekova2026AGNfields}, Version 1.2, doi:10.5281/zenodo.22045309. An interactive version of the resource is available at \url{https://izviekova-blip.github.io/agn-reference-fields/}.

\bibliographystyle{mnras}
\bibliography{references} % if your bibtex file is called example.bib

\clearpage
\appendix

\section{Additional diagnostics for IC~5287}
\label{app:ic5287_tests}
 
\subsection{Photometric and reverberation robustness}
\label{app:ic5287_phot_tests}
 
The principal chromatic-variability and optical--MIR reverberation results for IC~5287 are presented in Sections~\ref{subsec:results_variability} and \ref{subsec:results_reverberation}. Here we use additional optical colour combinations to test the robustness of the inferred BWB behaviour and examine the sensitivity of the MIR lag measurements to the accretion-disc subtraction, ICCF configuration, and individual WISE/NEOWISE epochs.

The optical BWB behaviour is recovered independently using colour combinations beyond the principal $g-r$ diagnostics shown in Fig.~\ref{fig:ic5287_colours}. The $r-i$ versus $r$ relation gives a York slope
\begin{equation}
m_{r-i}=0.691\pm0.108,
\end{equation}
with Pearson $r=0.594$ and $p<10^{-6}$. The corresponding $F_\nu(r)$--$F_\nu(i)$ flux--flux relation has an ODR slope
\begin{equation}
a_{ri}=0.767\pm0.123,
\end{equation}
corresponding to a variable-component colour of
\begin{equation}
(r-i)_{\rm var}=0.288~{\rm mag}.
\end{equation}
The $g-i$ colour--magnitude relation provides a consistent result, with a slope of $0.759\pm0.113$, confirming that the BWB trend is not specific to the principal $g-r$ colour combination.

In the adopted $F_\nu\propto\nu^{-\alpha}$ convention, the median
$r-i$ colour corresponds to an observed central-region broadband
spectral index
\begin{equation}
\alpha_{r-i}=1.753\pm0.030,
\end{equation}
with a brightness dependence
\begin{equation}
\frac{{\rm d}\alpha}{{\rm d}(-r)}
=
-3.079\pm0.483.
\end{equation}
Thus, the spectrum becomes flatter in $F_\nu$ as the source brightens, providing a spectral-index representation of the BWB trend independently recovered from the colour--magnitude relation.

The fiducial accretion-disc subtraction adopts $s_{\rm AD}=1/3$. Repeating the lag analysis with $s_{\rm AD}=0.1$ and $0$ changes the nominal $W1$ ICCF centroid by less than 0.02~d and the $W2$ centroid by less than approximately 0.6~d. These shifts are negligible compared with the statistical lag uncertainties, showing that the inferred reverberation scale is effectively insensitive to the assumed variable accretion-disc spectral contribution at the precision of the present data.

\begin{figure*}
\centering
\includegraphics[width=0.49\textwidth]
{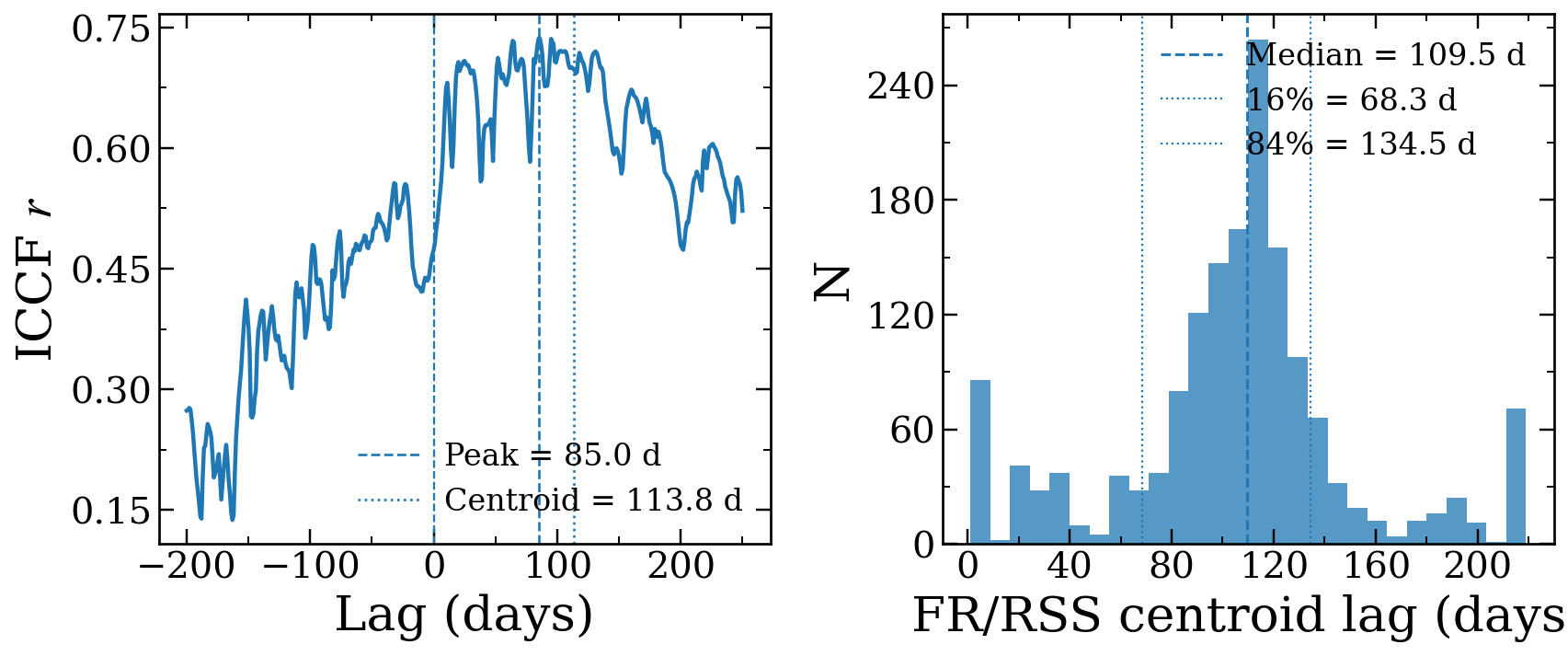}
\includegraphics[width=0.49\textwidth]
{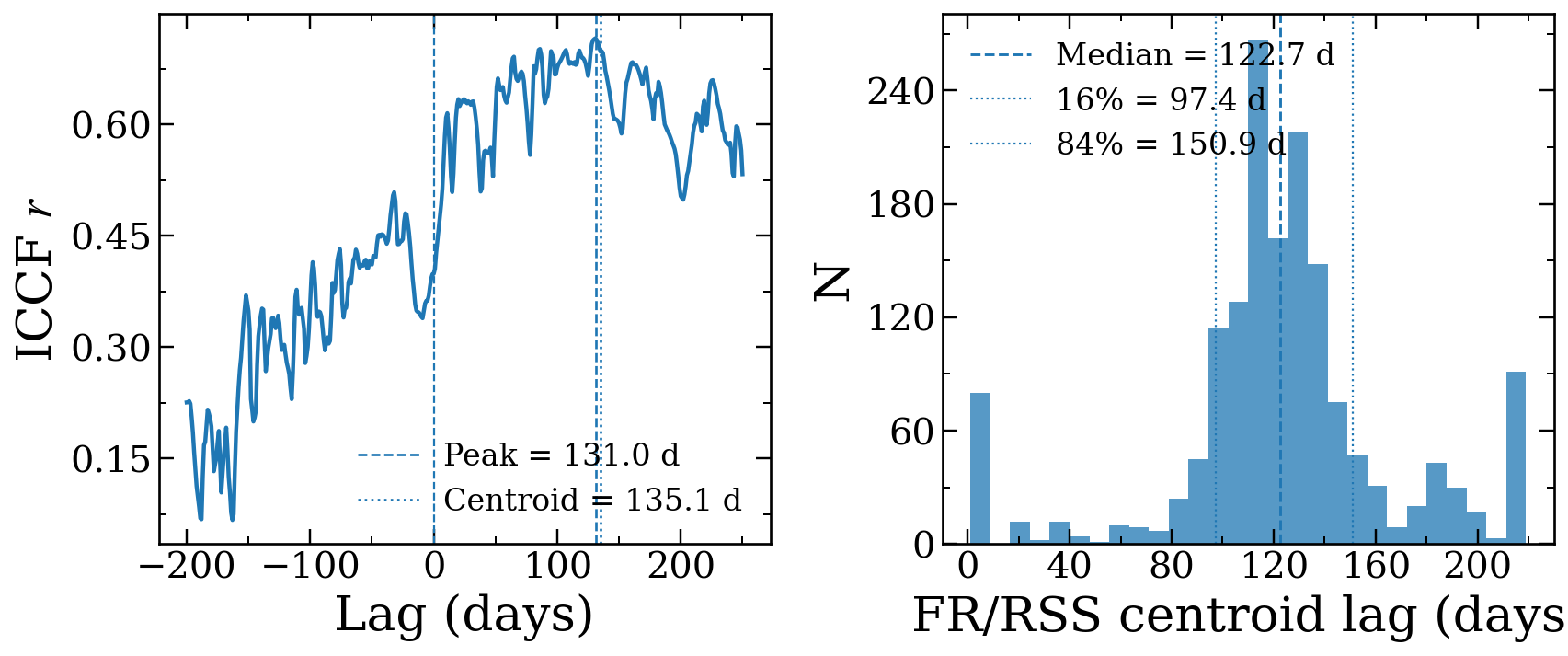}
\caption{ICCF and FR/RSS reverberation diagnostics for IC~5287 in $W1$ (left) and $W2$ (right). The panels show the fiducial bidirectional ICCFs and the corresponding FR/RSS centroid distributions after subtraction of the variable accretion-disc contribution using the fiducial $s_{\rm AD}=1/3$ prescription.}
\label{fig:ic5287_lags}
\end{figure*}

We additionally varied the ICCF centroid threshold and lag-search interval. Across these deterministic configurations, the observed-frame centroids span approximately 91.9--113.8~d in $W1$ and 95.4--135.6~d in $W2$, remaining within the range supported by the primary FR/RSS lag distributions. The fiducial ICCF profiles and corresponding FR/RSS centroid distributions are shown in Fig.~\ref{fig:ic5287_lags}.

Increasing the FR/RSS ensemble to 5000 realisations produced stable
centroid distributions: relative to the first 2000 realisations, the
median observed-frame lag changed by only 0.33~d in $W1$ and
0.17~d in $W2$.

Most leave-one-epoch-out tests remain close to the fiducial solution. The largest change occurs when the WISE/NEOWISE epoch near MJD~58644 is removed, shifting the preferred centroids to approximately 99~d in $W1$ and 78~d in $W2$. This epoch samples the steep rise of the dominant MIR variability feature and therefore has substantial leverage on the correlation centroid. Importantly, the optical--MIR response on a timescale of order $10^2$~d is retained, whereas the relative ordering of the $W1$ and $W2$ centroids is not. This behaviour reinforces the conclusion that the characteristic dust-reverberation scale of IC~5287 is robust, while the evidence for wavelength-dependent ordering remains tentative.

Overall, the alternative analysis configurations consistently recover an optical--MIR response on a timescale of order $10^2$~d. The dominant limitation is the sparse MIR sampling, which affects the precision of the response-weighted delay and, more strongly, the relative $W1$--$W2$ ordering.

\subsection{Broad H$\beta$ profile topology}
\label{app:ic5287_hbeta}

The broad H$\beta$ profile of IC~5287 is strongly non-Gaussian and shows a structured, double-peaked morphology. Because this structure can make the conventional FWHM sensitive to small changes in the fitted profile near the half-maximum level, we examined both the preferred profile complexity and the stability of alternative velocity-width measures.

Models containing one and two broad Gaussian components were compared
over the same spectral pixels using fixed observational uncertainties.
We define
\begin{equation}
{\rm BIC}
=
\chi^2+k\ln N,
\end{equation}
where $k$ is the number of free parameters and $N$ is the number of
fitted spectral pixels, and
\begin{equation}
\Delta{\rm BIC}
=
{\rm BIC}_{\rm 1G}
-
{\rm BIC}_{\rm 2G},
\end{equation}
such that positive values favour the two-component broad-line model.

Across 1000 Monte Carlo realisations, the median model preference is $\Delta{\rm BIC}=31.7$, with $\Delta{\rm BIC}>10$ in 95.6 per cent of the realisations. A single broad Gaussian is preferred in only approximately 0.3 per cent of the simulations. The structured broad-H$\beta$ morphology is therefore robust to the propagated spectral uncertainties.

The conventional FWHM is substantially less stable because the structured profile switches between two half-maximum topologies across the Monte Carlo realisations. In 69.8 per cent of the realisations, the inter-peak minimum remains
above the half-maximum level, so that the profile forms a single
connected region at half maximum and gives
\begin{equation}
{\rm FWHM}_{\rm high}
=
6889^{+351}_{-328}\ {\rm km\,s^{-1}}.
\end{equation}
In the remaining 30.2 per cent, the inter-peak minimum falls below the
half-maximum level. The conventional FWHM then follows a narrower
connected portion of the profile and decreases to
\begin{equation}
{\rm FWHM}_{\rm low}
=
3192^{+309}_{-390}\ {\rm km\,s^{-1}}.
\end{equation}
The resulting bimodality is therefore a consequence of applying the half-maximum definition to a structured profile: relatively small changes in the fitted flux near the inter-peak minimum can change the measured FWHM by more than a factor of two. The two FWHM modes should not be interpreted as two independent global BLR velocity scales.

To characterise the full broad-line profile, we calculate the
flux-weighted centroid and line dispersion of the continuum-subtracted
profile $P(v)$,
\begin{equation}
\bar v
=
\frac{\int vP(v)\,{\rm d}v}
     {\int P(v)\,{\rm d}v},
\qquad
\sigma_{\rm line}
=
\left[
\frac{\int (v-\bar v)^2P(v)\,{\rm d}v}
     {\int P(v)\,{\rm d}v}
\right]^{1/2}.
\end{equation}

We additionally calculate the cumulative-flux widths
$W_{50}=v_{75}-v_{25}$,
$W_{80}=v_{90}-v_{10}$, and
$W_{90}=v_{95}-v_{05}$, obtaining
\begin{equation}
\begin{aligned}
W_{50} &= 4114^{+168}_{-159}\ {\rm km\,s^{-1}},\\
W_{80} &= 6488^{+292}_{-254}\ {\rm km\,s^{-1}},\\
W_{90} &= 7747^{+369}_{-316}\ {\rm km\,s^{-1}}.
\end{aligned}
\end{equation}
Unlike the conventional FWHM, these integrated profile measures do not undergo a discrete change when the inter-peak minimum crosses the half-maximum level.

The centroid of the complete broad-H$\beta$ profile is consistent with
zero velocity offset within the Monte Carlo uncertainties. The line
dispersion is
\begin{equation}
\sigma_{\rm line}({\rm H}\beta)
=
2503^{+113}_{-98}\ {\rm km\,s^{-1}},
\end{equation}
and remains substantially more stable than the topology-dependent
conventional FWHM. The two representative FWHM topologies are
illustrated in Fig.~\ref{fig:ic5287_hbeta_topology}. We therefore adopt
$\sigma_{\rm line}({\rm H}\beta)$ as the primary broad-line velocity
measure entering the single-epoch BLR-scale prediction in
Section~\ref{subsec:results_spectroscopy}.

\subsection{Additional spectroscopic diagnostics}
\label{app:ic5287_line_diagnostics}
 
\subsubsection{Emission-line diagnostics}

The decomposed narrow-line ratios are
\begin{equation}
\log\frac{[\mathrm{O\,III}]}{\mathrm{H}\beta}\simeq1.107,
\quad
\log\frac{[\mathrm{N\,II}]}{\mathrm{H}\alpha}\simeq0.173,
\quad
\log\frac{[\mathrm{S\,II}]}{\mathrm{H}\alpha}\simeq-0.134,
\end{equation}
where only the decomposed narrow Balmer components are used. These ratios place the source securely in the AGN region of the standard BPT diagnostics.

The density-sensitive sulphur ratio is
\begin{equation}
\frac{F([\mathrm{S\,II}]\lambda6716)}
     {F([\mathrm{S\,II}]\lambda6731)}
=
1.202^{+0.108}_{-0.102},
\end{equation}
corresponding to
\begin{equation}
n_e
=
294^{+180}_{-141}\ {\rm cm^{-3}}
\end{equation}
for $T_e=10^4$~K.

The [O~III] profile contains an additional wing with
\begin{equation}
\Delta v_{\rm wing}
=
-103^{+31}_{-54}\ {\rm km\,s^{-1}},
\qquad
{\rm FWHM}_{\rm wing}
=
414^{+61}_{-64}\ {\rm km\,s^{-1}},
\end{equation}
and flux fraction
\begin{equation}
f_{\rm wing}
=
0.176^{+0.052}_{-0.043}.
\end{equation}
The complete [O~III] profile has $W_{80}=272^{+7}_{-6}$~km~s$^{-1}$, consistent with modest asymmetric NLR kinematics.

The optical Fe~II strength is
\begin{equation}
R_{\rm FeII}
=
\frac{F_{\rm FeII}(4434\mbox{--}4684\,{\rm \AA})}
     {F_{\mathrm{H}\beta,\rm broad}}
=
0.561^{+0.038}_{-0.033}.
\end{equation}

The detailed line diagnostics are consistent with the SDSS-epoch classification reported in Section~\ref{subsec:results_spectroscopy}.

\subsubsection{Stellar-continuum diagnostics}
\label{app:ic5287_stellar}

The robustness of the stellar velocity dispersion was tested with \textsc{pPXF} using alternative wavelength and mask configurations, with the wavelength-dependent SDSS instrumental resolution included explicitly. A redder-wavelength configuration gives approximately $125.4$~km~s$^{-1}$, differing from the fiducial value reported in Section~\ref{subsec:results_spectroscopy} by only a few km~s$^{-1}$.

The uncertainty in the AGN continuum fraction was assessed independently
with 150 pixel-noise continuum Monte Carlo realisations. The resulting
distribution has a median
$f_{\rm AGN,5100}=0.1441$ and a 16th--84th percentile range of
$0.1376$--$0.1509$. The accepted continuum-model configurations span
$0.1344$--$0.1456$. Adopting the outer limits of the statistical and
model-systematic intervals gives the conservative value reported in
the main text,
\begin{equation}
f_{\rm AGN,5100}
=
0.143^{+0.008}_{-0.009}.
\end{equation}

As an independent host diagnostic, the observed spectrum gives
\begin{equation}
D_n(4000)=1.676\pm0.012,
\end{equation}
whereas the isolated stellar component has $D_{n,\star}(4000)\simeq2.11$. The line-free AGN power law and AGN+Fe~II component give approximately 1.09 and 1.11, respectively.

The stellar fractions are approximately 68 per cent in the 3850--3950~\AA\ window and 80 per cent in the 4000--4100~\AA\ window. The intermediate observed break is consistent with dilution of the strong stellar 4000-\AA\ feature by the nuclear continuum.

\subsection{Internal reddening and BLR-scale systematics}
\label{app:ic5287_reddening}
 
The fiducial and attenuation-corrected continuum luminosities, their predicted H$\beta$ BLR scales, and the resulting dust-to-BLR ratios are reported in Section~\ref{subsec:results_scales}. Here we examine the internal consistency of the attenuation scenario and its sensitivity to the adopted broad-line velocity measure.

\subsubsection{Continuum-slope consistency}

The fiducial decomposed AGN continuum has $\beta_{\rm obs}=+0.25$ for $F_\lambda\propto\lambda^\beta$. Applying the BLR colour excess to the continuum produces the intrinsic slope reported in Section~\ref{subsec:results_scales}, which lies close to the standard long-wavelength thin-disc expectation
\begin{equation}
\beta=-7/3.
\end{equation}
Approximately 70 per cent of the propagated realisations lie within $\Delta\beta=0.3$ of this value. The agreement provides an internal consistency check on the attenuation-corrected scenario. The assumption
\begin{equation}
E(B-V)_{\rm cont}=E(B-V)_{\rm BLR}
\end{equation}
is nevertheless a modelling assumption and is not treated as an independent measurement of continuum extinction.

\subsubsection{Alternative FWHM-based estimates}

We additionally evaluated conventional FWHM-based estimators to assess the dependence of the BLR inference on the adopted velocity measure. For IC~5287, the structured H$\beta$ profile makes these estimates particularly sensitive to the half-maximum topology.

The \citet{VestergaardPeterson2006} H$\beta$ prescription gives
\begin{equation}
\log(M_{\rm BH}/M_\odot)
=
7.66^{+0.06}_{-0.64},
\end{equation}
where the asymmetric uncertainty reflects the connected and split FWHM branches. The broad-H$\alpha$ calibration of
\citet{GreeneHo2005} gives
\begin{equation}
\log(M_{\rm BH}/M_\odot)
=
7.366^{+0.014}_{-0.013}.
\end{equation}
These estimates are used only as auxiliary spectroscopic diagnostics; the main analysis adopts the independent literature black-hole mass.

Using the FWHM-based BLR relation of \citet{Woo2026}, the connected high-FWHM branch gives predicted lags of $5.43^{+4.26}_{-2.41}$~d for the fiducial continuum and $13.27^{+10.18}_{-5.79}$~d for the attenuation-corrected continuum. For the split low-FWHM branch, the corresponding values are $3.19^{+2.40}_{-1.37}$ and $8.50^{+6.18}_{-3.58}$~d.

The absolute values depend on the FWHM topology, but both branches show the same response to the continuum correction as the primary $\sigma_{\rm line}$-based calculation: increasing the inferred nuclear luminosity increases the predicted BLR scale.

\section{Additional diagnostics for Mrk~845}
\label{app:mrk845_tests}
  
\subsection{Photometric and reverberation robustness}
\label{app:mrk845_lag_tests}

The optical BWB behaviour of Mrk~845 is independently recovered from the $r-i$ colour. For the fiducial seeing threshold of $\leq3\arcsec$, the York slope is
\begin{equation}
\frac{{\rm d}(r-i)}{{\rm d}r}
=
0.329\pm0.027,
\end{equation}
with Pearson $r=0.672$ and $p<10^{-6}$. Restricting the sample to seeing $\leq2\arcsec$ gives $0.328\pm0.039$, demonstrating the stability of the relation against the adopted image-quality threshold.

The corresponding flux--flux relation gives
\begin{equation}
a_{ri}=1.040\pm0.042,
\end{equation}
corresponding to
$(r-i)_{\rm var}=-0.043$~mag and
$\alpha_{\rm var}^{r-i}=-0.19$.
Together with the principal $g-r$ diagnostics shown in Fig.~\ref{fig:mrk845_colours} and reported in Section~\ref{subsec:results_variability}, this supports a spectrally blue correlated optical component across the ZTF wavelength range.

In the adopted $F_\nu\propto\nu^{-\alpha}$ convention, the median $r-i$ colour corresponds to an observed central-region broadband spectral index
\begin{equation}
\alpha_{r-i}=1.353\pm0.018,
\end{equation}
with
\begin{equation}
\frac{{\rm d}\alpha}{{\rm d}(-r)}
=
-1.464\pm0.119.
\end{equation}
The negative derivative agrees with the $g-r$ spectral-index evolution and confirms BWB behaviour across both optical colour baselines.

The optical--MIR response was tested with top-hat and Gaussian transfer functions. The top-hat solution is strongly preferred by the BIC in both MIR bands, with a BIC weight of approximately 0.999 in $W1$ and effectively unity in $W2$. The corresponding best-fitting reconstruction is shown in Fig.~\ref{fig:mrk845_tf_reconstruction}.

The fitted response width is less stable than the characteristic lag and approaches the permitted width boundary in a substantial fraction of the realisations. We therefore do not use the fitted width as a direct measure of the geometrical thickness of the dust distribution.

Before enlargement of the uncertainties to include the full robustness set, the model-averaged bootstrap distributions give
\begin{align}
\tau_{\rm rest}(W1)
&=
156.8^{+8.9}_{-8.2}\ {\rm d},\\
\tau_{\rm rest}(W2)
&=
183.4^{+8.6}_{-12.0}\ {\rm d}.
\end{align}
These intervals describe the statistical lag distributions alone.

Additional tests varied the assumed accretion-disc contribution, treatment of long interpolation gaps, and individual MIR epochs. The same characteristic response scales are recovered, although the absolute lag is more sensitive to these choices than indicated by the bootstrap intervals alone. The conservative final delays and radii that encompass these robustness tests are reported in Section~\ref{subsec:results_reverberation}.

\begin{figure*}
\centering
\includegraphics[width=\textwidth]
{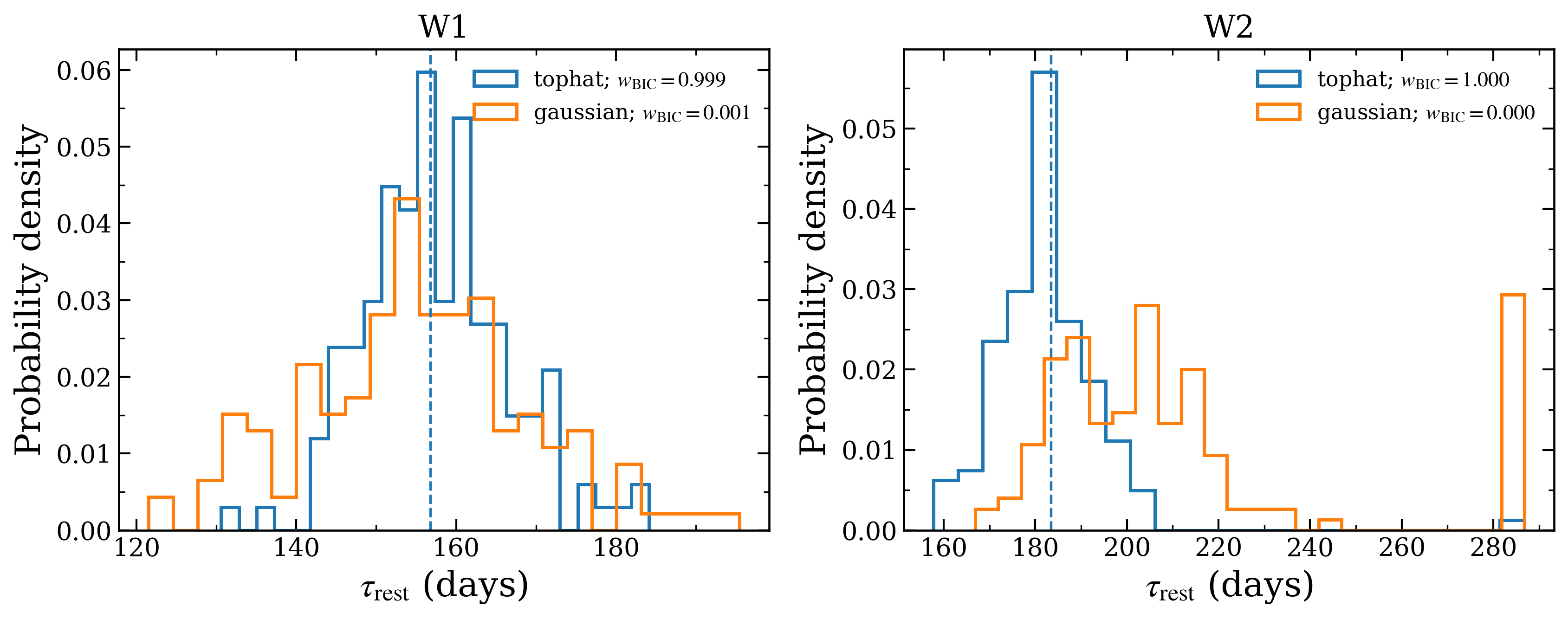}
\caption{Transfer-function bootstrap diagnostics for Mrk~845. The distributions compare the characteristic optical--MIR delays obtained with the tested response functions in $W1$ and $W2$. The characteristic delays are better constrained than the response widths. The final uncertainties reported in the main text additionally encompass the accepted robustness tests. }
\label{fig:mrk845_tf_bootstrap}
\end{figure*}

The wavelength ordering is retained in all paired leave-one-epoch-out, accretion-disc-subtraction, and interpolation-gap tests. The relative ordering of the $W1$ and $W2$ responses is therefore more stable than the fitted response width. The corresponding bootstrap lag distributions are shown in Fig.~\ref{fig:mrk845_tf_bootstrap}.

 \subsection{Spectroscopic diagnostics and robustness}
\label{app:mrk845_spectral_tests}
 
\subsubsection{Continuum and stellar-kinematic robustness}

The principal continuum quantities are reported in Section~\ref{subsec:results_spectroscopy}. Here we summarise the model-dependence and stellar-kinematic tests. The accepted continuum systematic set includes variations in the stellar kinematics and emission-line masks. Configurations in which nuisance parameters converge to imposed boundaries are excluded from the final systematic envelope. A fit without optical Fe~II is retained only as a model-dependence diagnostic.

An independent stellar-kinematic fit over 5100--6200~\AA\ gives approximately
\begin{equation}
\sigma_\star\simeq149.6\ {\rm km\,s^{-1}},
\end{equation}
consistent with both the fiducial value reported in the main text and the previously published value of approximately $139\pm17$~km~s$^{-1}$ \citep{Kompaniiets2025a}.

\subsubsection{Emission-line structure and model selection}

The preferred decomposition contains structured broad H$\beta$ and H$\alpha$ profiles together with a separate [O~III] wing. Monte Carlo model selection gives a two-Gaussian broad H$\beta$ model in 94.7 per cent of the realisations and a two-Gaussian H$\alpha$ model in all realisations.

For H$\beta$,
\begin{equation}
\Delta{\rm BIC}_{\rm H\beta}
=
31.7^{+14.9}_{-12.2},
\qquad
P(\Delta{\rm BIC}_{\rm H\beta}>10)=0.954.
\end{equation}
For H$\alpha$,
\begin{equation}
\Delta{\rm BIC}_{\rm H\alpha}
=
57.9^{+18.7}_{-17.1},
\end{equation}
with $\Delta{\rm BIC}>10$ in all realisations. The additional
[O~III] component is more strongly required,
\begin{equation}
\Delta{\rm BIC}_{[\mathrm{O\,III}]}
=
1262^{+85}_{-68},
\end{equation}
relative to the core-only model.

Beyond the H$\beta$ line dispersion adopted in the main analysis, the topology-marginalised profile gives
\begin{align}
{\rm FWHM}({\rm H}\beta)
&=
3413^{+253}_{-260}\ {\rm km\,s^{-1}},\\
W_{80}({\rm H}\beta)
&=
4602^{+149}_{-150}\ {\rm km\,s^{-1}}.
\end{align}
For H$\alpha$,
\begin{align}
\sigma_{\rm line}({\rm H}\alpha)
&=
1338^{+14}_{-16}\ {\rm km\,s^{-1}},\\
{\rm FWHM}({\rm H}\alpha)
&=
2676^{+76}_{-96}\ {\rm km\,s^{-1}},\\
W_{80}({\rm H}\alpha)
&=
3412^{+36}_{-41}\ {\rm km\,s^{-1}}.
\end{align}

\subsubsection{Narrow-line diagnostics}

\begin{figure*}
\centering
\includegraphics[width=0.92\textwidth]
{\detokenize{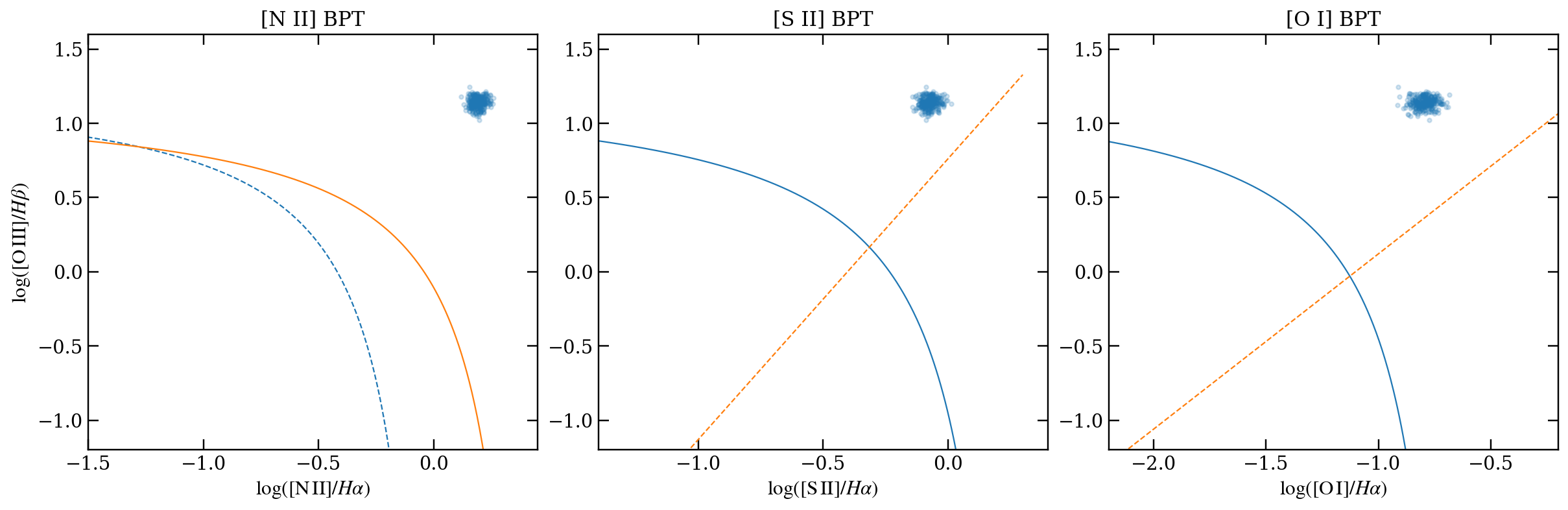}}
\caption{Monte Carlo narrow-line diagnostics for Mrk~845 in the standard BPT diagrams. The decomposed narrow-line spectrum remains in the AGN region of the [N~II] diagram and in the Seyfert regions of the [S~II] and [O~I] diagrams throughout the Monte Carlo ensemble.}
\label{fig:mrk845_bpt_mc}
\end{figure*}

The decomposed narrow-line ratios are
\begin{align}
\log\frac{[\mathrm{O\,III}]}{\mathrm{H}\beta}
&=
1.138^{+0.040}_{-0.034},\\
\log\frac{[\mathrm{N\,II}]}{\mathrm{H}\alpha}
&=
0.189^{+0.028}_{-0.022},\\
\log\frac{[\mathrm{S\,II}]}{\mathrm{H}\alpha}
&=
-0.074^{+0.027}_{-0.026},\\
\log\frac{[\mathrm{O\,I}]}{\mathrm{H}\alpha}
&=
-0.796^{+0.041}_{-0.038}.
\end{align}
All Monte Carlo realisations occupy the AGN region of the [N~II] diagram and the Seyfert regions of the [S~II] and [O~I] diagrams. The corresponding Monte Carlo BPT distributions are shown in Fig.~\ref{fig:mrk845_bpt_mc}.

The total-H$\beta$ to [O~III]$\lambda5007$ ratio is
\begin{equation}
\frac{
{\rm H}\beta_{\rm narrow}
+
{\rm H}\beta_{\rm broad}
}{
[{\rm O\,III}]\lambda5007
}
=
0.626\pm0.019.
\end{equation}
The resulting Seyfert subtype is reported in Section~\ref{subsec:results_spectroscopy}.

The [O~III] wing has
\begin{equation}
v_{\rm wing}
=
-165^{+11}_{-13}\ {\rm km\,s^{-1}},
\qquad
\sigma_{\rm wing}
=
554\pm13\ {\rm km\,s^{-1}}.
\end{equation}
The profile therefore contains a pronounced asymmetric ionised-gas component.

\section{Seasonal optical-timescale analysis of Mrk~845}
\label{app:mrk845_timescale_tests}

The ZTF $g$- and $r$-band light curves of Mrk~845 were divided into seven observing seasons using the same temporal intervals in both filters. A weighted linear trend was removed from each season, and a generalised Lomb--Scargle (GLS) periodogram was calculated over 20--110~d. The location of the strongest GLS peak defines the characteristic seasonal timescale, $P_{\rm GLS}$, used in the recurrence analysis.

\begin{figure}
\centering
\includegraphics[width=\columnwidth]
{\detokenize{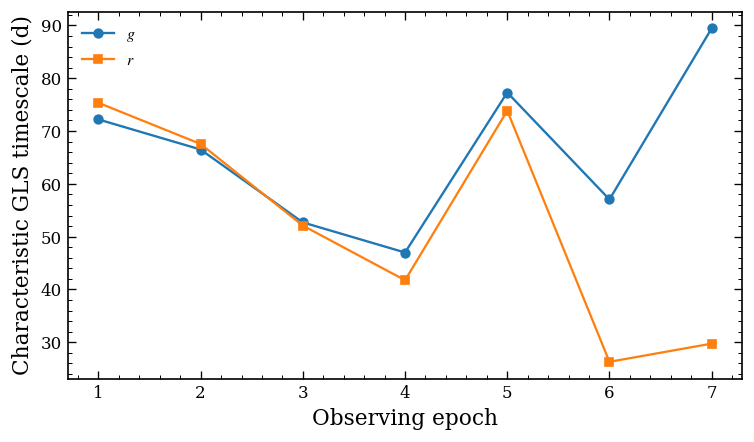}}
\caption{Characteristic seasonal GLS peak timescales of Mrk~845 in the ZTF $g$ and $r$ bands. The points show the locations of the strongest GLS peaks in each observing season. The seasonal values occupy a broad common timescale range; the statistical significance of their recurrence is evaluated independently with the DRW Monte Carlo analysis. }
\label{fig:mrk845_seasonal_timescales}
\end{figure}

The resulting seasonal GLS peak timescales are shown in Fig.~\ref{fig:mrk845_seasonal_timescales}. They span approximately 47--89~d in $g$ and 26--75~d in $r$. The two filters trace broadly similar characteristic timescales during the first five seasons, whereas the $r$-band estimates shift to shorter values during the final two seasons.

\begin{figure*}
\centering
\includegraphics[width=0.95\textwidth]
{\detokenize{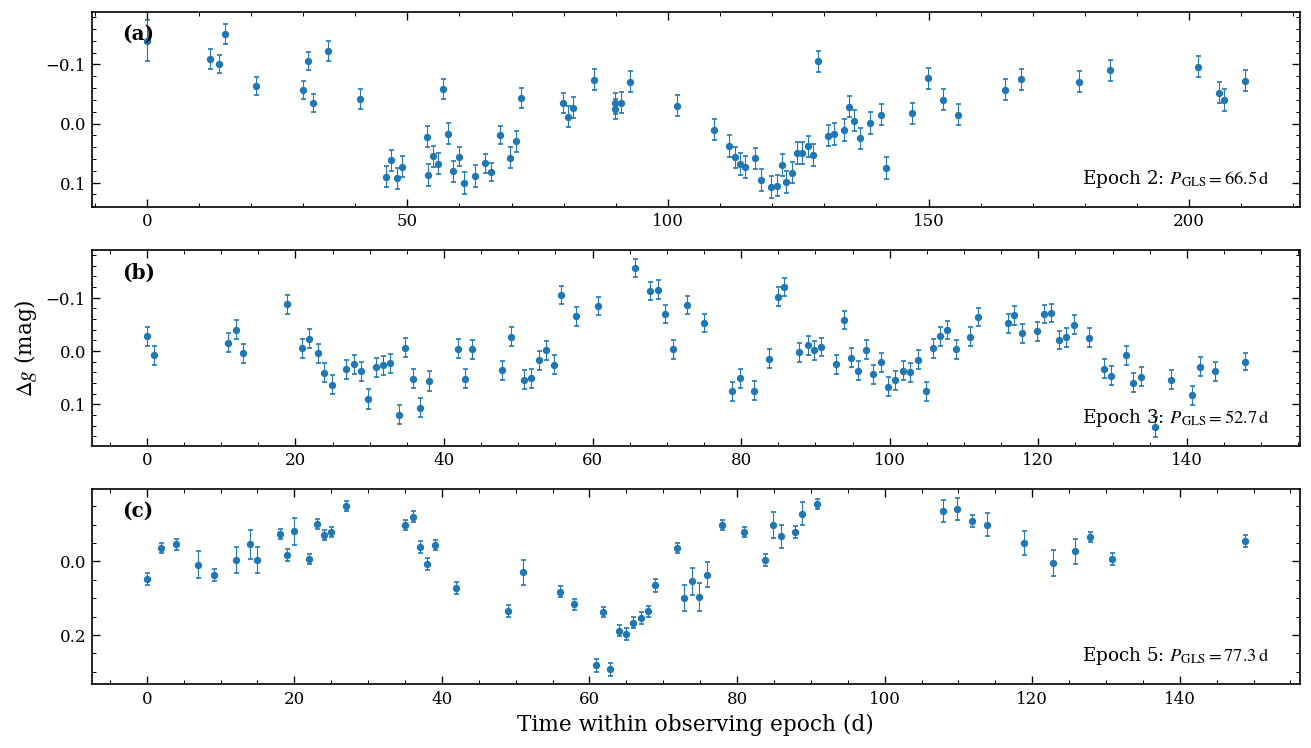}}
\caption{Detrended $g$-band light curves of three representative observing seasons of Mrk~845. The annotated values give the corresponding GLS characteristic timescales. Their recurrence across the observing seasons is quantified with the DRW Monte Carlo analysis. }
\label{fig:mrk845_representative_epochs}
\end{figure*}

Figure~\ref{fig:mrk845_representative_epochs} shows three representative detrended $g$-band seasons and illustrates the intra-seasonal structure associated with these characteristic timescales.

The significance of the recurrence was evaluated with parametric Monte Carlo simulations based on damped-random-walk (DRW) models fitted directly to the raw seasonal light curves,
\begin{equation}
m(t)=a+bt+X_{\rm DRW}(t),
\end{equation}
where the stochastic component was described by
\begin{equation}
k(\Delta t)=\sigma_{\rm DRW}^{2}
\exp\left(-|\Delta t|/\tau_{\rm DRW}\right).
\end{equation}
The likelihood included the reported photometric uncertainties and an additional fitted jitter term. We generated 30\,000 mock light curves at the observed sampling times and processed every realisation through the same detrending, GLS, and recurrence-search procedure as the observed data.

For each trial period band, the maximum GLS power was measured independently in every observing season and summed over the seven seasons. The recurrence statistic was obtained by maximising this sum over period-band centres and widths of 10--50~d within the full 20--110~d search interval. Applying the same scan to every DRW realisation incorporates the centre--width search into the empirical null distribution. The global probability was calculated as
\begin{equation}
p_{\rm global}
=
\frac{N_{\rm exc}+1}{N_{\rm MC}+1},
\end{equation}
where $N_{\rm exc}$ is the number of simulated statistics at least as large as the observed value and $N_{\rm MC}=30\,000$.

The resulting recurrence probabilities are $p_{\rm global}=6.0\times10^{-4}$ in $g$ and $p_{\rm global}=1.4\times10^{-2}$ in $r$. We also constructed a joint $g+r$ statistic by combining the maximum powers within the same trial period band through their geometric mean. Its null distribution was generated from a shared latent DRW process sampled at the actual $g$- and $r$-band observing times, with the band-dependent variability amplitudes and uncertainties preserved. This shared-process construction imposes strong cross-band stochastic coherence under the null and therefore provides a conservative joint test. The joint test gives $p_{\rm global}=8.7\times10^{-4}$.

\begin{figure*}
\centering
\includegraphics[width=0.94\textwidth]
{\detokenize{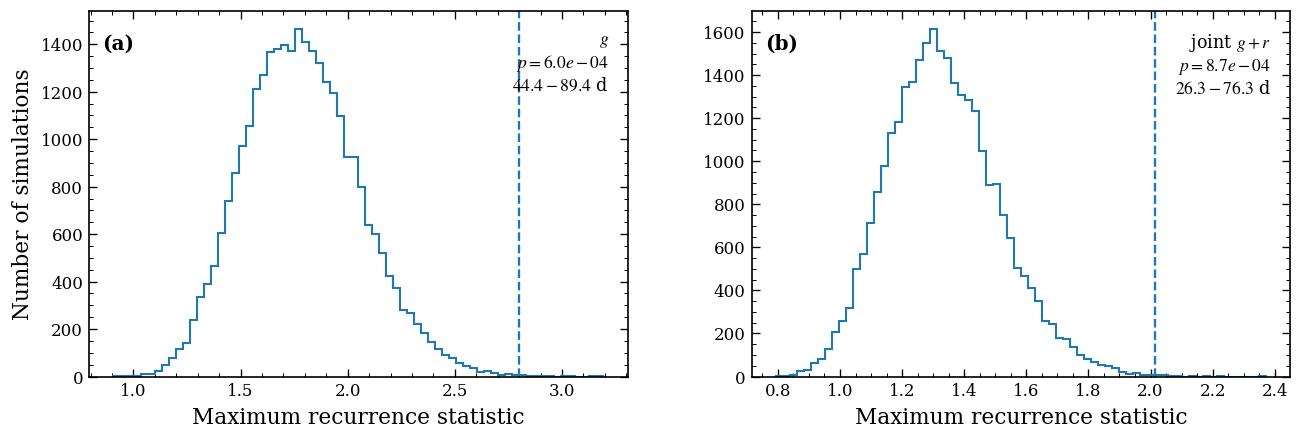}}
\caption{DRW Monte Carlo recurrence tests for Mrk~845. The histograms show the empirical null distributions of the recurrence statistic in the $g$ band (left) and for the joint $g+r$ analysis (right). The dashed lines mark the observed statistics. The global probabilities are $p_{\rm global}=6.0\times10^{-4}$ in $g$ and $p_{\rm global}=8.7\times10^{-4}$ for the joint statistic.}
\label{fig:mrk845_recurrence}
\end{figure*}

The corresponding empirical null distributions for the $g$-band and joint statistics are shown in Fig.~\ref{fig:mrk845_recurrence}.

The recurrence remains significant when individual observing seasons are removed. The leave-one-season-out probabilities span $4.7\times10^{-4}$--$2.8\times10^{-3}$ in $g$ and $5.0\times10^{-4}$--$3.1\times10^{-3}$ for the joint statistic, showing that the result is not driven by any single observing season.

As an independent test of temporal coherence, we compared linear-trend+DRW and linear-trend+DRW+SHO Gaussian-process models for all season--filter combinations. The BIC favours the DRW description throughout the sample. These results support recurrent structured stochastic variability on month-long timescales, with no evidence for a stable periodic component.

\section{Intraday optical variability}
\label{app:idv}

Dedicated high-cadence $V$-band monitoring with CAMELOT2 on the 0.82-m IAC80 telescope was used to test IC~5287 and Mrk~845 for variability on hour-long timescales. Aperture photometry was performed for each nucleus and a local comparison-star sequence. The comparison stars, their J2000 coordinates, APASS DR9 multiband photometry, and static finding charts are publicly available in the \textit{AGN Reference Fields} dataset \citep{Izviekova2026AGNfields}. Differential light curves were checked for seeing-dependent systematics before application of the power-enhanced $F$-test \citep{deDiego2014}, adopting $\alpha_{\rm sig}=0.01$.

The observing log is given in Table~\ref{tab:iac80_log}, and the final $F$-test statistics are summarised in Table~\ref{tab:idv_summary}.
\begin{table}
\centering
\caption{Log of the IAC80/CAMELOT2 high-cadence $V$-band observations. The UTC interval gives the start and end of each sequence; $N$ is the number of exposures retained for the final analysis.}
\label{tab:iac80_log}
\renewcommand{\arraystretch}{1.10}
\setlength{\tabcolsep}{6.2pt}
\small
\begin{tabular}{llcccc}
\hline
Object & Night & $t_{\rm exp}$ & UTC interval & Dur. & $N$ \\
       &       & (s)           &              & (h)  &     \\
\hline

IC~5287 &
Jun 30/Jul 1 &
90 &
02:10:38--05:10:55 &
3.00 &
101 \\

Mrk~845 &
May 20 &
60 &
21:01:55--01:16:35 &
4.25 &
125 \\

Mrk~845 &
May 21 &
60 &
21:06:32--01:04:32 &
3.97 &
116 \\

Mrk~845 &
May 22 &
70 &
20:55:28--00:58:00 &
4.04 &
112 \\
\hline
\end{tabular}
\end{table}
 
\subsection{IC~5287}
\label{app:idv_ic5287}
 
IC~5287 was analysed using three local comparison stars. Aperture radii of 5--13 pixels were tested for seeing-dependent systematics. The 5- and 7-pixel apertures produced strong correlations between target--reference differential magnitude and seeing ($\rho=-0.875$ and $-0.689$, respectively), whereas a 9-pixel radius ($2.90$~arcsec) minimized the dependence and was adopted for the final photometry.

The sky background was estimated from a sigma-clipped annulus separated from the source aperture by 80 pixels and having a width of 9 pixels. No Galactic-extinction correction was applied because it would produce only a constant offset in the differential light curve. For the adopted aperture, the final target--reference light curve shows no significant seeing dependence, with $r=0.149$ ($p=0.138$) and $\rho=0.110$ ($p=0.275$).

One miscentred exposure was rejected, leaving 101 frames. Comparison-star stability was tested with a leave-one-out procedure. Stars~1 and 2 were consistent with constant sources; star~3 showed one isolated anomalous measurement, whose inclusion or omission did not affect the variability classification.

\begin{figure}
\centering
\includegraphics[width=\columnwidth]
{\detokenize{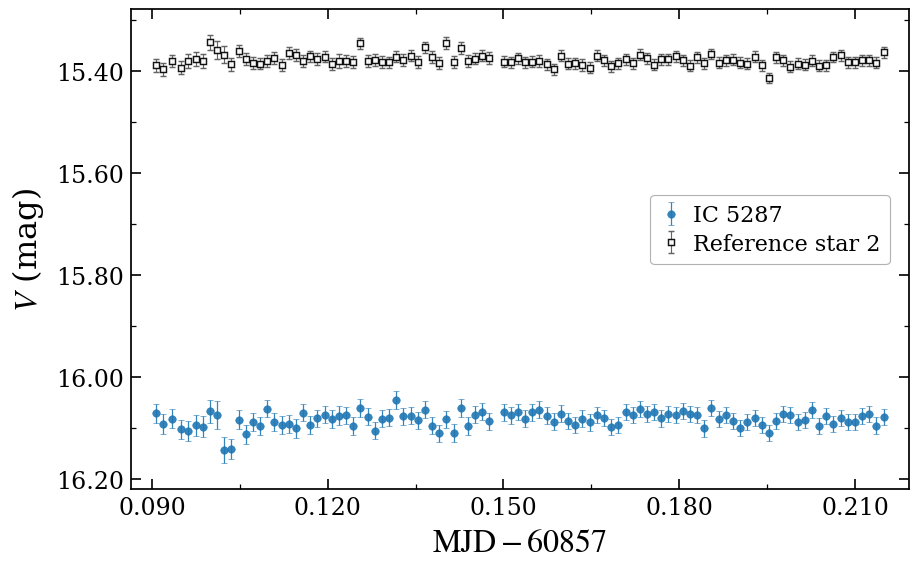}}
\caption{Observed $V$-band intraday light curve of IC~5287 obtained with IAC80/CAMELOT2 on 2025 June 30/July 1. Blue filled circles show the target and open squares show comparison star~2 as a photometric stability control. A 9-pixel ($2.90$~arcsec) aperture was used. Error bars represent propagated $1\sigma$ photometric uncertainties. }
\label{fig:ic5287_idv}
\end{figure}

Star~1 was adopted as the primary statistical reference. The power-enhanced $F$-test classifies the sequence as non-variable at $\alpha_{\rm sig}=0.01$. The result is unchanged when stars~2 or 3 are used as the primary reference or when the isolated anomalous measurement of star~3 is omitted. The resulting differential light curve is shown in Fig.~\ref{fig:ic5287_idv}.

\begin{table}
\centering
\caption{Power-enhanced $F$-test results for the IAC80 monitoring. All sequences are classified as non-variable at $\alpha_{\rm sig}=0.01$. }
\label{tab:idv_summary}
\small
\setlength{\tabcolsep}{6.0pt}
\renewcommand{\arraystretch}{1.05}

\begin{tabular}{llccc}
\hline
Object & Night & $F_{\rm enh}$ & $F_{\rm crit}$ & $p$ \\
\hline
IC~5287 & Jun 30/Jul 1 & 0.898 & 1.481 & 0.724 \\
Mrk~845 & May 20 & 0.873 & 1.390 & 0.813 \\
Mrk~845 & May 21 & 0.901 & 1.407 & 0.742 \\
Mrk~845 & May 22 & 0.837 & 1.415 & 0.865 \\
\hline
\end{tabular}
\end{table}

 \subsection{Mrk~845}
\label{app:idv_mrk845}
 
Mrk~845 was analysed using four local comparison stars. Because the source is spatially extended at the CAMELOT2 resolution, the same aperture was used on all three nights. Tests over radii of 8--12 pixels led to the adoption of a 9-pixel ($2.90$~arcsec) radius.

The same sky-background prescription as for IC~5287 was used. The comparison stars were tested with a leave-one-out procedure; star~3 was adopted as the primary statistical reference on May 20 and star~4 on May 21 and 22. APASS DR9 lists $\sigma_V=0.000$~mag for star~4; the variability analysis uses the differential-photometry uncertainties and the empirical stability of the comparison-star sequence.

For the adopted aperture, the Pearson/Spearman correlations between target--reference differential magnitude and seeing are $(0.046,\,0.100)$, $(-0.175,\,-0.205)$, and $(0.032,\,0.066)$ for May 20, 21, and 22, respectively. The weak May 21 rank correlation ($p_{\rm Spearman}=0.027$) does not affect the variability classification.

\begin{figure}
\centering
\includegraphics[width=\columnwidth]
{\detokenize{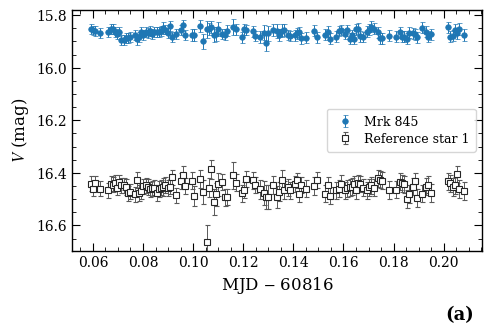}}
\vspace{1.5mm}
\includegraphics[width=\columnwidth]
{\detokenize{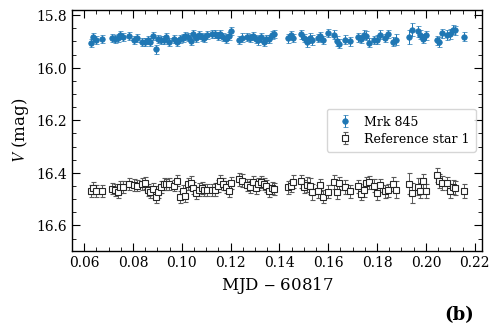}}
\vspace{1.5mm}
\includegraphics[width=\columnwidth]
{\detokenize{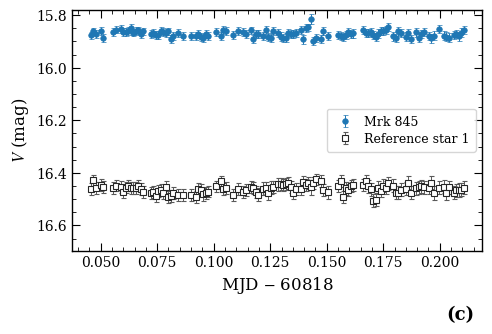}}
\caption{Observed $V$-band intraday light curves of Mrk~845 obtained with IAC80/CAMELOT2 on 2025 May 20 (a), May 21 (b), and May 22 (c). Blue filled circles show Mrk~845 and open squares show comparison star~1 as a photometric stability control. A common 9-pixel ($2.90$~arcsec) aperture was used. Error bars represent propagated $1\sigma$ photometric uncertainties. Stars~3 and 4 were the primary statistical references on May 20 and May 21--22, respectively. }
\label{fig:mrk845_idv}
\end{figure}

The power-enhanced $F$-test yields an NV classification for all three Mrk~845 sequences at $\alpha_{\rm sig}=0.01$. The corresponding differential light curves are shown in Fig.~\ref{fig:mrk845_idv}. The IAC80 monitoring therefore yields no statistically significant intraday variability in either IC~5287 or Mrk~845 during the sampled intervals.
 
\label{lastpage}

\end{document}